\documentclass[prd,aps,twocolumn,floats,floatfix,nofootinbib]{revtex4-1}

\usepackage{amssymb,bm}
\usepackage{graphicx}
\usepackage{dcolumn}
\usepackage{bm}
\usepackage{graphics}
\usepackage{slashed}
\usepackage{natbib}
\usepackage{amsmath}
\usepackage{url}
\usepackage[usenames,dvipsnames,svgnames,table]{xcolor}
\usepackage{color}
\usepackage{xcolor}

\definecolor{color1}{RGB}{191, 0, 255}

\newcommand{\sign}{{\rm sgn}}

\begin{document}

\title{GRMHD large eddy simulations with gradient subgrid-scale model}

\author{
Daniele Vigan\`o$^{1,2,3,4}$,
Ricard Aguilera-Miret$^{1,2}$,
Federico Carrasco$^{1,2,5}$,
Borja Mi\~nano$^{2}$,
Carlos Palenzuela$^{1,2}$
}

\affiliation{${^1}$Departament  de  F\'{\i}sica,  Universitat  de  les  Illes  Balears  and  Institut  d'Estudis
Espacials  de  Catalunya,  Palma  de  Mallorca,  Baleares  E-07122,  Spain\\
$^2$Institut Aplicacions Computationals (IAC3),  Universitat  de  les  Illes  Balears,  Palma  de  Mallorca,  Baleares  E-07122,  Spain\\
$^3$Institut d'Estudis Espacials de Catalunya (IEEC), 08034 Barcelona, Spain\\
$^4$Institute of Space Sciences (ICE, CSIC), 08193 Barcelona, Spain\\
$^5$Max Planck Institute for Gravitational Physics (Albert Einstein Institute), 14476 Potsdam, Germany}
 
\begin{abstract}

The detection of binary neutron star mergers represents one of the most important astrophysical discoveries of the recent years. Due to the extreme matter and gravity conditions and the rich dynamics developed, it becomes a tremendous challenge to accurately simulate numerically all the scales present during the collision. Here we present how to study such systems by using large eddy simulations with a self-consistent subgrid-scale gradient model, that we generalized to the special relativistic case in a previous work and now extend to the general relativistic case. Adapted from nonrelativistic scenarios, the so-called gradient model allows to capture part of the effects of the hidden dynamics on the resolved scales, by means of a physically-agnostic, mathematically-based Taylor expansion of the nonlinear terms in the conservative evolution equations' fluxes.
We assess the validity of this approach in bounding-box simulations of the magnetic Kelvin-Helmholtz instability.
Several resolutions and a broad range of scenarios are considered in order to carefully test the performance of the model under three crucial aspects: (i) highly curved backgrounds, (ii) jumps on the fluid density profiles and (iii) strong shocks. The results suggest our extension of the gradient subgrid-scale model to general relativistic magnetohydrodynamics is a promising approach for studying binary neutron stars mergers, and potentially to other relevant astrophysical scenarios.

\end{abstract}

\maketitle

\section{Introduction}

One of the most important scientific discoveries of the last decade has been the first detection of gravitational waves from the coalescence of a binary neutron star (BNS) system by the LIGO/Virgo facilities~\cite{abbott17a,abbott17b}. The event GW170817 was followed by a broadband electromagnetic signal: a gamma-ray burst (GRB) after less than two seconds (GRB170817A)~\cite{abbott17d}, a thermal infrared/optical radiation consistent with a kilonova \cite{abbott17c, metzger17} starting from hours after, and X-ray \cite{davanzo18} and radio \cite{dobie18} afterglow, the latter been visible for many months.
The observed signals are consistent with a BNS merging into a short-lived hypermassive neutron star (HMNS), which collapses to a black hole (BH) surrounded by an accretion disk. In particular, short GRBs are believed to be the high-energy manifestation of a relativistic jet, mainly powered by the rotational energy of the BH and the surrounding post-merger disk through the accretion process~\cite{lee07}.
This single multi-messenger detection already led to important breakthroughs in our understanding of the universe, providing clues to fundamental physics such as the properties of matter at nuclear densities, emission mechanisms in extreme plasma conditions, and stringent tests of General Relativity (GR) (e.g. \cite{burns19}) among others.

Magnetic fields are believed to play a fundamental role on the dynamics of these systems, especially in launching the magnetically dominated jets. However, the formation channels of BNS systems and their magnetic fields are quite uncertain. Observationally, we can infer the surface magnetic fields in old pulsars belonging to binary systems to be in the range of $10^8-10^{11}$ G \cite{tauris17}. Such magnetic fields are expected to be modified by local, small-scale amplification of the intensity by several orders of magnitude during the merger through several processes. The first one acting is the Kelvin-Helmholtz instability (KHI), which develops at the shear layer that forms between the external layers of the colliding NSs. The KHI acts only in the first few milliseconds of the merger and leads to a fast growth of small-scale magnetic fields, with a cut-off wavelength of the order of the size of the discontinuity layer. The simulations of BNS mergers with the current finest numerical resolutions, approaching $\Delta \sim 10$ m (where $\Delta$ is the grid size) \cite{kiuchi15,kiuchi18}, are still not able to fully capture the KHI: although they provide some notable amplification of small-scale magnetic structures, there is no yet sign of numerical convergence, meaning that dynamically relevant scales are still unresolved.

After the quick growth of the magnetic fields due to the KHI, there are two mechanisms associated to the differentially rotating HMNS which are expected to play an important role on longer timescales of 10-100 ms: magnetic winding, which creates and linearly amplifies a toroidal magnetic fields starting from a poloidal component, and the magneto-rotational instability (MRI). For the latter, the wavelength of the fastest growing modes is proportional to the magnetic fields. Some recent works manage to fully resolve the MRI by modelling only the post-merging phase and setting an initial large-scale strong magnetic field of $10^{15}-10^{16}$ G (e.g., \cite{kiuchi18,ciolfi19}).
During and after these processes, the resulting magnetic fields intensity and configuration, the energy/helicity kinetic/magnetic spectra, and the formation of coherent structures are important features that affect the jet properties, the mass ejecta and the properties and fate of the newly formed HMNS \cite{ciolfi18}. Note however that the seed magnetic field configuration assumed in the simulations is usually a large-scale dipolar field, which artificially favours the appearance and amplification of an ordered magnetic field and the jet formation, but is inconsistent with the expected outcome turbulent spectra emerging from the KHI, which distributes magnetic energy over all scales. Therefore, there is great interest in understanding how these magnetic field amplification on the small scales can be transferred to the large ones, which might happen due to the combination of dynamo effect and inverse cascade of magnetic energy.

In brief, although the main magnetohydrodynamics (MHD) during and after the merger is qualitatively understood, only very accurate and demanding numerical simulations can model these processes and foresee their possible outcome. Such simulations are extremely difficult due to the intrinsic nonlinearity of the equations involved and on the wide range of spatial and temporal scales of MHD processes. To even further complicate the problem, strong gravity effects are important in NSs, implying that Einstein equations need also to be solved. Due to the unfeasible computational resources required, no study has so far consistently simulated all the phases of the magnetic dynamics described above with the required accuracy to address these issues.
At contrast with the purely hydrodynamic case, the finite resolution hampers the ability of reproducing an efficient dynamo mechanism and therefore affects the large-scale dynamics, since MHD turbulence tends to distribute the magnetic energy (amplified at small scales) across the whole spectrum.
In the absence of computationally viable direct numerical simulations with a spatial resolution able to capture all the relevant scales, the dynamo processes have been simulated in GRMHD by including additional terms in the induction equation \cite{bucciantini13,giacomazzo15,palenzuela15}. While these approaches provide an effective growth of the magnetic fields, they have fundamental issues: they do not converge to the continuum version of the equations for very high resolution, and they need a tuning and switching on/off “by hand” of the extra terms, which is arbitrary and virtually impossible to be calibrated on physical grounds.

In order to find an inexpensive approximation able to capture the turbulent regime and allow back-scattering and dynamo mechanisms, in this work we borrow the explicit Large Eddy Simulation (LES) approach\footnote{Technically, all BNS mergers simulations performed so far qualify --by definition-- as implicit LESs, since they do not solve all the relevant small scales, and the only effects of the unresolved scales are provided implicitly (unwillingly and with no control) by the intrinsic dissipation/dispersion of the numerical scheme. Here we instead focus on explicit LESs.}, common in different areas like engineering and plasma physics (see, e.g., \cite{lu16}), but scarcely utilized in relativistic MHD.
In an explicit LES approach, the evolution equations for the resolved fields are supplemented with explicit sub-grid-scale (SGS) terms which effectively model the small scales \cite{zhiyin15}. So far, to the best of our knowledge, only one work has so far introduced an SGS model into a GR simulation, limiting the study to the non-magnetic hydrodynamic case \cite{radice17}. The author took into account the dissipative effects of magnetic fields on the post-merger differentially rotating fluid by implementing a purely viscous SGS model \cite{smagorinsky63}, originally designed for incompressible hydrodynamics. Note that, by construction, that particular SGS model only allows for an effective inclusion of dissipative mechanisms, and it cannot reproduce neither the relevant transfer of kinetic to magnetic energy (dynamo), nor the redistribution of the latter over a broad range of scales (back-scattering), relevant in BNS mergers and in most turbulent MHD scenarios. Therefore, in order to capture further effects arising from the SGS dynamics, one may need more sophisticated SGS models.

In this paper, we extend our previous SGS gradient models for nonrelativistic \cite{vigano19b} and special relativistic \cite{carrasco19} MHD, to full GR(MHD). The main advantage of the gradient SGS model (originally designed for incompressible fluids \cite{leonard75,muller02a}) is that it does not rely on any physical assumption: it is based on the mathematical properties of the particular set of equations describing the dynamics. In this paper, we implement the proposed SGS model in a code using high-order accurate finite-difference numerical methods. We validate the approach by focusing on 3D bounding box simulations of the KHI. We explore  a broad range of problem's parameters, including the most relevant aspects for the BNS scenario, namely: (i) spacetime curvature, (ii) sharp gradients in the initial fluid density profile, and (iii) strong shocks in the velocity field.  
We perform two different kind of tests: \textit{a-priori}, comparing the residuals of the subfilter-scale (SFS) tensors from a high resolution run --without the SGS model-- to the SGS tensors proposed by the model; and \textit{a-posteriori}, in which the high-resolution run is now compared against a low-resolution run evolved using the SGS terms.
The results of these tests support the present GR extension of our SGS model, and suggest it could be a promising tool to tackle the KHI developed during the merger in BNS. The model is general, and relevant also for the subsequent post-merger phase, as well as applicable to many other astrophysical scenarios.      

This article is organized as follows. We summarize the formalism in \S\ref{sec:les} and the numerical methods and analysis in \S\ref{sec:methods}. We then show the applications in bounding box KHI simulations in \S\ref{sec:box3d}, where we compare the results between a high-resolution, accurate simulation, with an explicit LES at a lower resolution. Finally, conclusions are drawn in \S\ref{sec:discussion}.

\section{Large Eddy Simulations of GRMHD}\label{sec:les}

LES techniques are classically employed in scenarios where not all the dynamically relevant scales can be resolved simultaneously: therefore, the large scales are simulated, while the small ones are only modeled by the inclusion of SGS terms, effectively capturing the effects of these small scales on the resolved ones. 
In any numerical evolution problem, the effect of discretization over a finite-resolution grid can be associated to an effective filtering of the continuum equations, with the filter size being of the order of the grid-cell.
In other words, the information inside the cell is partially lost by the discrete representation of the numerical solution, regarded in this context as a weighted average of the values within the cell.
If one formally applies a filtering operator to the set of evolution equations, new terms will appear due to the non-commutativity of the filtering operator acting on the nonlinear terms of the equations.
These new, a-priori unknown terms, represent the SFS residuals. The aim of an explicit LES approach is to model them, by using the available resolved fields, in order to close the system of evolution equations. For completeness, we will briefly summarize some of the results presented in our previous works~\cite{vigano19b,carrasco19}. Let us start from a generic set of conservation laws,
\begin{equation}
\partial_t C^a + \partial_k F^{ka} (P) = 0~, \label{eq:conservation_law}
\end{equation}
where $C^a$ is a set of {\em conserved} evolved variables and $F^{ka}(P)$ is the flux along the direction $k$ of the field $C^a$, which can be expressed in terms of the {\em primitive} fields $P^a$. 
The connection between conserved and primitive fields can be formally written as\footnote{Although the relation $f^a$ between conserved and primitive fields is known explicitly, the inverse function $g^a$ is not analytical in relativistic ideal MHD, where it needs to be solved numerically.}
\begin{equation*}
	C^a = f^a (P) ~,~~~
	P^a :=  (f^{-1})^{a}(C) \equiv g^{a}(C)~.
\end{equation*}
For any given variable $C^a ({\bf x},t)$, the filtering operator separates the resolved part ${\overline C}^a ({\bf x},t)$ from the unresolved one $C^{a'}({\bf x},t)$ as,
\begin{equation}
    C^{a}({\bf x},t) = {\overline C}^{a}({\bf x},t) + C^{a'}({\bf x},t)~,
\end{equation}
with
\begin{equation}
{\overline C}^{a}({\bf x},t) :=  \int_{-\infty}^{\infty} G({\bf x}-{\bf x'}) C^{a}({\bf x'},t) d^3 {\bf x'}~,
\end{equation}
being $G({\bf x}-{\bf x'})$ the filter's kernel function, with compact support.
The filtered version of the system in eq.~\eqref{eq:conservation_law} then reads
\begin{equation}
\partial_t \overline{C}^a + \partial_k F^{ka} (\widetilde{P}) = \partial_k \overline{\tau}^{ka}~,
\end{equation}
where the SFS residuals (related to the fluxes) are defined by
\begin{equation}\label{residual-F}
\overline{\tau}^{ka} :=  F^{ka} (\widetilde{P}) -  \overline{F^{ka}(P)} ~~.
\end{equation}
Here we have defined $\widetilde{P}^a :=  g^a (\overline{C})$, which can be directly computed from the evolved $\overline{C}^a$, as opposed to $\overline{P}^a \equiv \overline{g^{a}(C)}$, which are the ones directly appearing after the filtering, but not known in a finite-size-grid simulation due to the unavoidable SFS information loss.
The second term in eq.~\eqref{residual-F} is, therefore, the piece that needs to be conveniently modeled by SGS terms, which may depend only on the known available fields, i.e. $\widetilde{P}^a$ and ${\overline C}^a$.   

After a detailed study in nonrelativistic KHI simulations \cite{vigano19b}, for which different available SGS models were compared, we have focused our attention on the gradient model. 
At contrast with other SGS models, this model is physically agnostic, in the sense that it does not depend on guessing which kind of physical effects dominate at small scales (dissipation at a certain scale, inverse cascade, etc.) in a specific scenario. 
Additionally, it can be applied to any problem and set of equations, since it relies solely on the mathematical structure of the system, extrapolating, cell by cell and term by term, the SGS dynamics by using the information contained in the gradients of the fields involved. In that sense, it resembles high-order reconstruction methods, effectively capturing small-scale effects with no significant extra computational cost.

Following Ref.~\cite{carrasco19}, the gradient SGS model was applied to the general system \eqref{eq:conservation_law} above, yielding to a formal general solution, first order-accurate in the gradient expansion, namely
\begin{equation}\label{tau-alt} 
\overline{\tau}^{ka} \backsimeq -\xi \, \nabla  \frac{dF^{ka}}{d\overline{C}^b}  \cdot \nabla \overline{C}^b~, 
\end{equation}
where $\xi:= \Delta^2 / 24$ for a flat Minkowski spacetime and $\Delta$ represents the numerical grid spacing. Note that the gradient model (as most SGS models) is designed to converge to the continuous solution in the high resolution limit, since the SGS terms are proportional to $\Delta^2$.

Below, we apply this formulation to the GRMHD equations. After summarizing our formulation of the Einstein and the GRMHD equations, we extend our previous special relativistic work~\cite{carrasco19} to the full GR setting, under the assumption that the metric components are smooth and slowly varying, as compared to the matter fields.

\subsection{Einstein equations}

The spacetime geometry is described by the Einstein equations, which can be extended in a convenient way by using the covariant conformal Z4 formulation (CCZ4)~\cite{alic12}, namely
\begin{eqnarray}
   R_{ab} + \nabla_a Z_b + && \nabla_b Z_a = 
   8\pi \, \left( T_{ab} - \frac{1}{2}g_{ab} \,{\rm tr} T \right)\nonumber \\
  && + \kappa_{z} \, \left(  n_a Z_b + n_b Z_a - g_{ab} n^c Z_c \right)\,,
\label{Z4cov}
\end{eqnarray}
where $R_{ab}$ is the Ricci tensor associated to the spacetime metric $g_{ab}$ and $T_{ab}$ is the total stress-energy tensor with trace ${\rm tr} T \equiv g^{ab} T_{ab}$.  The $Z_{a}$ four-vector measures deviations from Einstein's solutions~\cite{bona03,bona04} (i.e., those satisfying the constraints). Notice that suitable damping terms, proportional to the parameter $\kappa_{z}>0$, have been included in order to enforce the dynamical decay of the constraint violations associated with $Z_{a}$~\cite{bezares17}.

The covariant equations can be written as an evolution formalism by performing the $3+1$ decomposition~\cite{bonabook}. The line element can be decomposed as
\begin{equation}
ds^2 = - \alpha^2 \, dt^2 + \gamma_{ij} \bigl( dx^i + \beta^i dt \bigr) \bigl( dx^j + \beta^j dt \bigr)~, 
\label{3+1decom}  
\end{equation}
where $\alpha$ is the lapse function, $\beta^{i}$ is the shift vector, and $\gamma_{ij}$ is the induced metric on each spatial foliation. In this foliation, the normal to the hypersurfaces 
can be defined as $n_{a}=(-\alpha,0)$ and the extrinsic curvature as $K_{ij} \equiv  -\frac{1}{2}\mathcal{L}_{n}\gamma_{ij}$,  where $\mathcal{L}_{n}$ is the Lie derivative along  $n^{a}$. 

Subsequently, a conformal decomposition is applied to the evolved fields, which consists of performing a conformal transformation to the metric
and the extrinsic curvature, i.e., $\gamma_{ij}=\tilde{\gamma}_{ij}/\chi$ with $\chi$ being the conformal factor and $\tilde{\gamma}_{ij}$ having a unit determinant, and $K_{ij}$ into a trace $trK$ and a trace-less part $\tilde{A}_{ij}$. This transformation leads to two new constraints, which can also be enforced dynamically by including additional damping terms to the evolution equations~\cite{bezares17}. The final set of evolution fields, together with the gauge conditions setting the choice of coordinates, can be found in~\cite{palenzuela18}.

\subsection{GRMHD equations} 

Here we consider the usual stress-energy tensor associated to a magnetized, non-viscous and perfectly conducting fluid (e.g., \cite{palenzuela15}), and we use the usual relativistic units. The set of primitives variables describing this perfect magnetized fluid is given by $P^a = \left\lbrace \rho, v^i , \epsilon, B^i \right\rbrace$:  $\rho$ is the rest-mass density of the fluid, $\epsilon$ its specific internal energy, $v^{i}$ is the velocity vector and $B^i$ is the magnetic field.
However, evolution equations are expressed in terms of a different set of fields, known commonly as the (densitized) conserved ones, given by  $C^a = \sqrt{\gamma} \left\lbrace D, S^i , U, B^i \right\rbrace$. The relation between these two sets of fields is given by
\begin{eqnarray}
	D &=& \rho W ~, \\
	S^i &=& (h W^2 + B^2 ) v^i - (v\cdot B) B^i ~,\\
	U &=& h W^2 - p + B^2 - \frac{1}{2} \left[ (v\cdot B)^2 + \frac{B^2}{W^2} \right]  ~,
\end{eqnarray}
where $W= (1-v^2 )^{-1/2}$ is the Lorentz factor, the pressure $p$ is defined through an Equation of State (EoS) and the enthalpy is defined as $h= \rho (1 + \epsilon) + p$.
The evolution equations for conserved fields can be written as follows:
\begin{eqnarray}
&& \partial_t (\sqrt{\gamma} D) 
+ \partial_k [- \beta^k \sqrt{\gamma} D + \alpha \sqrt{\gamma} N^k] = 0 ~ ~,
\label{evol_D} \\
&& \partial_t (\sqrt{\gamma}  S_i) + \partial_k [- \beta^k \sqrt{\gamma} {S}_i + \alpha \sqrt{\gamma} {T^{k}}_i] = \sqrt{\gamma} {R^S}_i
\label{evol_S} ~, \\
&& \partial_t (\sqrt{\gamma} U) + \partial_k [- \beta^k \sqrt{\gamma} {U} + \alpha  \sqrt{\gamma} S^{k}]  = \sqrt{\gamma} R^U  
\label{evol_U} ~,\\
&& \partial_t (\sqrt{\gamma} B^i) + \partial_k [\sqrt{\gamma}(- \beta^k {B}^i  +  \beta^i {B}^k) 
\nonumber \\
&& \quad\quad\quad\quad + \alpha \sqrt{\gamma} ({\gamma}^{ki}{\phi} + M^{ki})] = \sqrt{\gamma} {R_B}^i ~,
\label{evol_B} \\
&& \partial_t (\sqrt{\gamma}{\phi}) + \partial_k [- \beta^k \sqrt{\gamma} {\phi} + \alpha\, c_h^2 \sqrt{\gamma}{B}^k] = \sqrt{\gamma} R^{\phi} ~,
\label{evol_phi} 
\end{eqnarray}
where we have introduced a new field $\phi$ associated to the hyperbolic divergence cleaning of the solenoidal constraint (as in previous works, e.g. \cite{palenzuela18}). The fluxes have been written explicitly in connection with their special relativistic counterparts:
\begin{eqnarray}
N^k &=& {v}^k {D} ~,\\
M^{ki} &=& B^{i} v^{k} - B^{k} v^{i} ~,  \\
T^{ki} &=& h W^2 v^k v^i - E^k E^i - B^k B^i  + \gamma^{ki} \left[ p+ \frac{1}{2}(E^2 + B^2 ) \right]  \nonumber \\
 &=& \frac{1}{2} \left({v}^i {S}^j + {v}^j {S}^i \right) +  \gamma^{ij} {p} - \frac{1}{{W}^2} \bigg( {B}^i {B}^j - \frac{1}{2} \gamma^{ij} {B}^2  \bigg) \nonumber \\    
&-& \frac{1}{2} ({B}^k {v}_k) \bigg[ {B}^i {v}^j + {B}^j {v}^i - \gamma^{ij} ({B}^m {v}_m)  \bigg]~.
\end{eqnarray}
Note that the ideal MHD condition $E^i = -\epsilon^{ijk} \, v_j \, B_k $ allows to write easily the previous fluxes and sources in terms of evolved and primitive fields. For numerical accuracy reasons, in our simulations we actually recombine two equations in order to evolve the conserved field $\sqrt{\gamma}(U - D)$ instead of $\sqrt{\gamma}U$.

The source terms, written already as a function of conformal variables, are
\begin{eqnarray}
R^U &=& \frac{\alpha}{\chi} {T}^{ij} {\tilde A}_{ij} + \frac{\alpha}{3} tr{T}\, trK
- {S}^j \partial_j \alpha ~, \nonumber \\
R^S_i &=& \frac{\alpha}{2 \chi} \left({T}^{jk} {\partial_{i}} {\tilde \gamma}_{jk} - tr{T} \partial_i \chi \right) + {S}_j \partial_i \beta^j - U \partial_i \alpha ~, \nonumber \\
R_B^i &=&  {\phi} \left[ - \alpha \chi {\tilde \Gamma}^i + {\tilde \gamma}^{ki} \left(- \frac{\alpha}{2} \partial_k \chi + \chi \partial_k \alpha\right) \right] ~,   \nonumber \\
R^{\phi} &=& - \alpha\,  {\phi}\, trK + c_h^2 {B}^k (\partial_k \alpha)  -\alpha \kappa {\phi} ~, \nonumber
\end{eqnarray}
where $tr {T} = \gamma_{jk} {T}^{jk} =  {\tilde \gamma}_{jk} {T}^{jk}/\chi$ and $ {\tilde \Gamma}^i = {\tilde \gamma}^{ij} {\tilde \gamma}^{kl} \partial_l {\tilde \gamma}_{jk}$.

\subsection{Filtered GRMHD equations} 

We now apply the filtering operator to the equations. First, notice that it is transparent to the partial derivatives, so it is straightforward to calculate the filtered MHD equations. Second, as it was mentioned at the beginning of the section, during a numerical simulation one can not calculate the filtered nonlinear flux terms $\{ \overline{N}^k, \overline{{T}^{k}}_i, \overline{S}^k \}$, but instead $\{ \widetilde{N}^k, \widetilde{T}^{k}_i, \widetilde{S}^k \} := \{ {N}^k(\widetilde{P}), {{T}^{k}}_i(\widetilde{P}), {S}^k(\widetilde{P}) \}$, that is, the fluxes calculated with the primitive fields $\widetilde{P}^a$, obtained from the inversion of the filtered conserved fields $\overline{C}^a$. The difference results in SFS terms, which are corrections to the fluxes. Third, we assume that SFS terms arising from the metric functions are negligible compared to the ones arising from the turbulent fluids (see a more detailed discussion at the end of this section).
Under these considerations, the filtered version of the GRMHD system can be written as: 

\begin{eqnarray}
&& \partial_t (\sqrt{\gamma} \overline{D}) + \partial_k [- \beta^k \sqrt{\gamma} \overline{D} + \alpha \sqrt{\gamma} (\widetilde{N}^k -\overline{\tau}^{k}_{N})] = 0 ~,
\label{evol_D_sgs} \nonumber \\
&& \partial_t (\sqrt{\gamma}  \overline{S}_i) + \partial_k [- \beta^k \sqrt{\gamma} \overline{{S}}_i + \alpha \sqrt{\gamma}( \widetilde{T}^{k}_i - \gamma_{ij} \overline{\tau}^{jk}_{T})] = \sqrt{\gamma} \overline{{R^S}}_i~,
\label{evol_S_sgs}  \nonumber \\
&& \partial_t (\sqrt{\gamma} \overline{U}) + \partial_k [- \beta^k \sqrt{\gamma} \overline{{U}} + \alpha  \sqrt{\gamma} (\widetilde{S}^{k} - \overline{\tau}^{k}_{S})]  = \sqrt{\gamma} \overline{R^U}  ~,
\label{evol_U_sgs}  \nonumber \\
&& \partial_t (\sqrt{\gamma} \overline{B}^i) + \partial_k [\sqrt{\gamma}(- \beta^k \overline{{B}}^i  +  \beta^i \overline{{B}}^k) 
\nonumber \\
&& \quad\quad\quad\quad + \alpha \sqrt{\gamma} ({\gamma}^{ki}\overline{{\phi}} + \widetilde{M}^{ki} - \overline{\tau}^{ki}_{M} )] = \sqrt{\gamma} \overline{{R_B}}^i ~,
\label{evol_B_sgs} \\
&& \partial_t (\sqrt{\gamma}\overline{{\phi}}) + \partial_k [- \beta^k \sqrt{\gamma} \overline{{\phi}} + \alpha\, c_h^2 \sqrt{\gamma}\overline{{B}}^k] = \sqrt{\gamma} \overline{R^{\phi}}~, 
\label{evol_phi_sgs} 
\end{eqnarray}
where we have introduced the SFS corrections inside each flux. The SGS gradient terms approximating such corrections can be written considering the non-trivial nonlinear dependencies of fluxes on the conserved variables, and following the general rule \eqref{tau-alt}:
\begin{eqnarray}
\tau^{k}_{N}  = -~{\cal C}~\xi \, H_{N}^k ~~&,&~~
\tau^{ki}_{T} = -~{\cal C}~\xi \, H_{T}^{ki} ~~, \nonumber \\
\tau^{k}_{S}  = 0  ~~&,&~~
\tau^{ki}_{M} = -~{\cal C}~\xi \, H_{M}^{ki} ~~. \label{eq:sgs_gradient}
\end{eqnarray}
where $\xi= \gamma^{1/3} \Delta^2/24$ scales appropriately to be consistent with the volume element of the spacetime. The parameter ${\cal C}$ is theoretically meant to be one, but, as we will see below and as shown in \cite{vigano19b}, the optimum value (i.e., the one best mimicking the SFS residuals in a LES) can differ depending on the numerical methods employed, being in general ${\cal C}\gtrsim 1$. Therefore, it is advisable to leave ${\cal C}$ as the only free parameter of the SGS model.

The set of the $H$ tensors, after some algebraic manipulations, can be obtained explicitly by computing the set of equations written below, in the order in which they appear, where the quantities $\widetilde{\Psi}$ denote auxiliary fields which are used to simplify the implementation:
\begin{widetext}
	\begin{eqnarray}
	\widetilde{\Psi}_{v}^k &=& \frac{2}{\widetilde{\Theta}} \left\lbrace \nabla (\widetilde{v}\cdot \overline{B}) \cdot \nabla \overline{B}^k  - \nabla \widetilde{\Theta} \cdot \nabla \widetilde{v}^k   
	+ \frac{\overline{B}^k}{\widetilde{\mathcal{E}}} \left[  \widetilde{\Theta} \nabla \overline{B}^j \cdot \nabla \widetilde{v}_j + \overline{B}_j \nabla \overline{B}^j \cdot \nabla (\widetilde{v}\cdot \overline{B}) - \overline{B}^j \nabla \widetilde{v}_j \cdot \nabla \widetilde{\Theta} \right]  \label{hTauv}  \right\rbrace ~, \nonumber  \\
	\widetilde{\Psi}^{ki}_{M} &=& \frac{4}{\widetilde{\Theta}} \left[  \widetilde{\Theta} \, \nabla \overline{B}^{[i} \cdot \nabla \widetilde{v}^{k]} +  \overline{B}^{[i} \nabla \overline{B}^{k]} \cdot \nabla (\widetilde{v}\cdot \overline{B}) - \overline{B}^{[i} \nabla \widetilde{v}^{k]} \cdot \nabla \widetilde{\Theta} \right] ~,
	\nonumber \\
	\widetilde{\Psi}_{\Theta} &=&  \frac{\widetilde{\Theta}}{\widetilde{\Theta} -\widetilde{E}^2} \left\lbrace \nabla \overline{B}_{j} \cdot \nabla \overline{B}^{j} - \nabla \widetilde{E}_{j} \cdot \nabla \widetilde{E}^{j} - \overline{B}_{[i}\widetilde{v}_{k]} \, \widetilde{\Psi}^{ki}_{M} \right\rbrace ~~,~~
	\widetilde{\Psi}_{A}  = \widetilde{W}^2 \left( \widetilde{p} \, \frac{d\widetilde{p}}{d\widetilde{\epsilon}} + \widetilde{\rho}^2 \, \frac{d\widetilde{p}}{d\widetilde{\rho}} \right) ~, \nonumber \\
	H_{\rm p} &=&  \frac{\widetilde{\mathcal{E}} \, \widetilde{W}^2 ({\widetilde{\Theta}- \widetilde{E}^2 })}{(\widetilde{\rho} \, \widetilde{\mathcal{E}} - \widetilde{\Psi}_{A})(\widetilde{\Theta} - \widetilde{E}^2 ) \widetilde{W}^2 + \widetilde{\Psi}_{A} \, \widetilde{\Theta}} \left\lbrace \widetilde{\rho} \left( \nabla \frac{d\widetilde{p}}{d\widetilde{\rho}} \cdot \nabla \widetilde{\rho} + \nabla \frac{d\widetilde{p}}{d\widetilde{\epsilon}} \cdot \nabla \widetilde{\epsilon} \right)  - 2 \frac{d\widetilde{p}}{d\widetilde{\epsilon}} \, \nabla \widetilde{\rho} \cdot \nabla \widetilde{\epsilon}  \right. \nonumber\\
	&-&  \left.  \left(\widetilde{\mathcal{E}} \frac{d\widetilde{p}}{d\widetilde{\epsilon}} - \widetilde{\Psi}_{A}\right) \left[ \frac{\widetilde{W}^2}{4} \nabla \widetilde{W}^{-2} \cdot \nabla \widetilde{W}^{-2} + \nabla \widetilde{W}^{-2} \cdot \nabla (\ln \widetilde{\rho}) \right]  -  \frac{2}{\widetilde{W}^2}\frac{d\widetilde{p}}{d\widetilde{\epsilon}} \left[   \nabla \overline{B}_j \cdot \nabla \overline{B}^j -  \widetilde{W}^{4} \nabla \widetilde{W}^{-2} \cdot \nabla \widetilde{h} \right]  \right. \label{tau_p} \\
	&-&  \left.  \left(\widetilde{\mathcal{E}} \frac{d\widetilde{p}}{d\widetilde{\epsilon}} + \widetilde{\Psi}_{A}\right) \left[ \widetilde{v}_j \widetilde{\Psi}_{v}^j +  \nabla \widetilde{v}_{j} \cdot \nabla \widetilde{v}^{j} + \widetilde{W}^2 \, \nabla \widetilde{W}^{-2} \cdot \nabla \widetilde{W}^{-2} \right]  +   \frac{\widetilde{\Psi}_{\Theta}}{\widetilde{\mathcal{E}} \widetilde{\Theta}} \left[ \left(\widetilde{\mathcal{E}} \frac{d\widetilde{p}}{d\widetilde{\epsilon}} + \widetilde{\Psi}_{A}\right)(\widetilde{\Theta}- \widetilde{E}^2 ) - \frac{\widetilde{\Psi}_{A} \, \widetilde{\Theta}}{\widetilde{W}^2}   \right]   \right\rbrace ~, \nonumber \\
	H_{\Theta} &=& \widetilde{\Psi}_{\Theta} + \frac{\widetilde{\Theta}}{\widetilde{\Theta} -\widetilde{E}^2} H_p ~, \label{tau_Theta} \\ 
	H_{v}^k &:=& \widetilde{\Psi}_{v}^k - \left( \widetilde{v}^k + \frac{\widetilde{v}\cdot \overline{B}}{\widetilde{\mathcal{E}}} \overline{B}^k \right)  \frac{H_{\Theta}}{\widetilde{\Theta}} ~, \label{Tauv} \\
    H^{k}_{N} &=&  2 \, \nabla \overline{D} \cdot \nabla \widetilde{v}^k + \overline{D} \, H^{k}_v ~~,~~
    H^{ki}_{M} =  2 \overline{B}^{[i} H_{v}^{k]} + 4 \, \nabla \overline{B}^{[i} \cdot \nabla \widetilde{v}^{k]} ~~\rightarrow~~ 
	H_{E}^i = \frac{1}{2} \epsilon^{i}_{\phantom ijk } H_{M}^{jk} ~,
	\label{HNME} \\
    H^{ki}_{T} &=& 2 \left[ \nabla \widetilde{\mathcal{E}} \cdot \nabla (\widetilde{v}^k \widetilde{v}^i ) + \widetilde{\mathcal{E}} \left( \widetilde{v}^{(k} H_{v}^{i)} +  \nabla \widetilde{v}^{k} \cdot \nabla \widetilde{v}^{i} \right)  +  \widetilde{v}^k \widetilde{v}^i H_{p} \right] 
     - 2\left[ \nabla \overline{B}^{k} \cdot \nabla \overline{B}^{i} + \nabla \widetilde{E}^{k} \cdot \nabla \widetilde{E}^{i} + \widetilde{E}^{(k} H_{E}^{i)}   \right]  \nonumber \\
    &+& (\gamma^{ki} - \widetilde{v}^k \widetilde{v}^i)  \left[ H_p + \nabla \overline{B}_{j} \cdot \nabla \overline{B}^{j} + \nabla \widetilde{E}_{j} \cdot \nabla \widetilde{E}^{j} + \widetilde{E}_{j} H_{E}^{j} \right]~, \label{HT}     
	\end{eqnarray}
\end{widetext}
where we have used the shortcuts $\mathcal{E} = h W^2$, 
$\Theta = \mathcal{E} + B^2$, and the two gradients $\nabla$ (on each term) symbolize spatial partial derivatives $\partial_i$ (and $\partial_j$), with ``$\cdot$'' indicating contraction among them with the spatial metric $\gamma^{ij}$.
Notice that, in order to include generic EoS, we also need the derivatives of pressure with respect to other thermodynamic fields. Therefore, the set of auxiliary fields in our implementation is given by $P^a = \left\lbrace \rho, \epsilon, p , v^i , B^i, \phi, dp/d\rho, dp/d\epsilon \right\rbrace $.

The expressions above extend our previous calculations~ \cite{carrasco19}, which assumed a flat Minkowski spacetime with $\sqrt{\gamma}=\alpha = 1$ and $\beta^i = 0$, to the general
relativistic case. Here, the metric is instead arbitrary, but is considered transparent to the gradient operators, which
implicitly assumes that the metric is much smoother than the fluid fields. This approximation has two consequences: (i) we neglect any SFS correction to the Einstein equations, (ii) we deal with the SFS terms of the MHD equations (within the $3+1$ formalism) essentially as in the nonrelativistic case, accommodating the curved spacetime elements (lapse function, shift vector and spatial metric) which appear in the flux terms. In other words, we neglect the metric gradients in the fluxes $\overline{C}^a = \sqrt{\gamma} \left\lbrace \overline{D}, \overline{S}^i , \overline{U}, \overline{B}^i, \overline{\phi} \right\rbrace$, and we neglect the SFS terms in the source terms.

This is justified by the fact that the fields prone to turbulence, to which the evolution of resolved scales is sensitive, are the MHD ones, and not the metric. In other words, derivatives of the metric components are negligible compared to the ones of velocity, magnetic field, density and pressure (and their combinations). These are expected to be reasonable approximations when MHD turbulence is developed, even in the presence of strong curvature like in a BNS merger, because the large variations of the turbulent fields at small scales would be the dominant effects we want to capture with the SGS terms.

Finally, we also neglect the SFS terms in the solenoidal magnetic field constraint equation, since it is designed to keep $\phi$ under control but does not represent a meaningful physical turbulent field.

\section{Numerical setup and analysis}\label{sec:methods}

\subsection{Numerical methods}

The generation of our computational code is performed by using the in-house-built and publicly available platform {\it Simflowny} \cite{arbona18}, developed to solve generic partial differential equations. It has a modular design which allows an independent and user-friendly implementation of the physical model, problem setup, and numerical schemes (time advance, space discretization and reconstruction methods). It generates efficient codes for the {\tt SAMRAI} infrastructure \cite{hornung02,gunney16}, which allows an excellent scaling of parallel computation performances over thousands of processors.

The code has been deeply tested for different scenarios \cite{palenzuela18,vigano19,2020arXiv200207554L}, including basic tests of MHD and GR. As in our previous works \cite{vigano19b,carrasco19}, we use the Method of Lines, which allows to address separately the time and the space discretization. The space-time evolution equations are discretized in space by using centered finite-difference, fourth-order-accurate operators for the derivatives, and sixth-order Kreiss-Oliger dissipation to filter
the high-frequency modes unresolved in our grids.  For the fluid part, we employ High-Resolution Shock-Capturing (HRSC) methods \cite{toro97} to deal with the possible appearance of shocks and to take advantage of the existence of weak solutions in the equations. The fluxes at the cell interfaces are calculated by combining the Lax-Friedrich flux splitting formula \cite{shu98}, with the fifth-order, monotonicity-preserving reconstruction method MP5 (which showed better performances for turbulent fluids, compared to other methods \cite{palenzuela18}). The SGS terms are computed with fourth-order centered finite-difference operators. The time integration of the resulting semi-discrete equations is performed by using a fourth-order Runge-Kutta scheme, which ensures the stability and convergence of the solution for a small enough time step $\Delta t \leq 0.4 ~\Delta$.  
We employ a divergence-cleaning scheme, which ensures that the deviations from the constraint, quantified by the dimensionless ratio between the volume-integrated values of the L2-norm of $\Delta (\vec{\nabla}\cdot\vec{B})$ and the magnetic energy, is always (typically) less $\lesssim 10^{-4}$. For further details about the divergence-cleaning formalism and numerical results, see our previous works \cite{palenzuela18,vigano19}.

\subsection{A-priori tests}

A first assessment of the performance of the SGS model, including its correct implementation, is the \textit{a-priori} test. In order to do it, one has to run simulations with a certain grid-spacing $\Delta$ and consider a snapshot at a given time. Then, one has to apply a spatial filter to all the conserved fields; the simplest recipe is to use a simple average groups of $S_f^3$ cells, where we define $S_f$ as the filter factor. The information lost in the filtering process contained in the scales $\in [\Delta , S_f\Delta]$ (therefore, it is not the entire loss of information, which is by definition unknown), for a given nonlinear term, is the difference between the filter of the entire term and the nonlinear combination of the filtered values of the conserved fields. Such partial SFS information
can be numerically evaluated as $\overline{\tau}(\vec{x}_f)$ at each location $\vec{x}_f$ of the filtered grid, and compared to the corresponding SGS model $\tau(\vec{x}_f,{\cal C})$. The performance is then evaluated by the classical Pearson correlation coefficient ${\cal P}$ between the SFS and SGS quantities, together with the best-fit value of the free parameter ${\cal C}$ in the SGS terms, as defined by eq.~(\ref{eq:sgs_gradient}). This best-fit value, which theoretically should be close to 1, can be easily calculated by using the classical analytical formula for a linear regression fit. The calculation of ${\cal P}$ and ${\cal C}$ is done for each SFS/SGS term. In our previous works \cite{vigano19b,carrasco19}, the gradient model showed very good performance (${\cal P}\gtrsim 0.8$) for all terms (better than other SGS models available in the nonrelativistic case), and degraded to ${\cal P}\lesssim 0.5$ only for quite large filter factors $S_f \gtrsim 8$.
Below we will show similar results for relativistic KHI in a curved background.

\subsection{A-posteriori tests}

Although informative, the \textit{a-priori} tests only provide how well a SGS model fits the instantaneous SFS residuals between $\Delta$ and $\Delta_f$ in a snapshot, over a certain range of parameters. The second class of tests is instead more challenging, since it considers the feedback of the included SGS terms, accumulated over time. It consists in comparing a low-resolution+SGS simulation with reference results. The latter are ideally represented by an analytical solution, or by numerical solutions to which simulations show numerical convergence. However, most of the times they are not available, so one has to take a high-resolution simulation as a reference.

The comparison between simulations can be done at different levels. A first one is represented by the volume-integrated quantities, especially the magnetic energy in our case. At a more detailed level, very informative whenever there is turbulence, it is illustrative to compute also the radially-averaged spectrum \cite{durran17,vigano19b}. For a given field $f$ defined in a periodic box of side $L$, we use common {\tt python} functions to calculate its discrete fast Fourier transform $\hat{f}(\vec{k}) = \Sigma_{\vec{x}} f(\vec{x}) e^{-i \vec{k}\cdot\vec{x}}$, where the sum is performed over the $N^3$ points equally spaced in each direction, with $k_j = n~\Delta k$, where $\Delta k = \frac{2\pi}{L}$ and $n\in [0,N/2]$ is an integer. Then, we calculate the solid-angle-averaged values $4\pi < k^2 |f|^2 >_{k}$ over the radial bins in the Fourier space, centered at $k=\{n~\Delta k\}$, which represent the power density per unit of radial wave-number. This defines the kinetic and magnetic spectra,
\begin{eqnarray}
&& \mathcal{E}_k(k) = \frac{L^3 4\pi}{(2\pi)^3N^6} <k^2|\widehat{\sqrt{\rho}\vec{v}}|^2(\vec{k})>_{k}~, \nonumber\\
&& \mathcal{E}_m(k) = \frac{L^3 4\pi }{(2\pi)^3N^6}<k^2|\hat{\vec{B}}|^2(\vec{k})>_{k}~, ~\label{eq:spectra}
\end{eqnarray}
that we define for simplicity as in the nonrelativistic case.

Below, we will do a-posteriori tests with different resolutions, activating the SGS terms with different values of the free parameter ${\cal C}$.

\section{Numerical simulations}\label{sec:box3d}

We test our approach in bounding box simulations, analogous and generalized compared to what can be found in the literature for the nonrelativistic cases \cite{obergaulinger10,beckwith11,vigano19b}, and in our previous special relativistic work  \cite{carrasco19}. Here, we consider a fixed metric, setting to zero the time evolution of the spacetime fields, evolving only the MHD part. This allows us to test the effect of a curved spacetime on the turbulence, but without the additional complication of a dynamical metric. In all the following simulations, we evolve the system at least to the saturation time, after which the solution approaches a stationary regime, with the turbulence completely developed and slowly decaying due to numerical dissipation.

\subsection{2D single-vortex KHI}

Since turbulent MHD is intrinsically a nonlinear and complex phenomenon, our first benchmark test considers only a simplified case with the creation of a single vortex via KHI in a flat metric, which is just the relativistic analogous of previous studies~\cite{obergaulinger10,vigano19b}.

We set a rectangular domain in the $x-y$ plane, centered in the origin, with side length $L_x=1$, $L_y=2$, with a number of points $(N,2N)$ equally spaced in the two directions. Boundary conditions are periodic in $x$-direction and radiative in the $y$-direction. The initial conditions consist of $\rho=1$ and $p=0.24$ everywhere, with a shear layer of thickness $a_l$ representing the initial discontinuity at $y=0$, and the following velocity and magnetic fields:
\begin{eqnarray}
&& v_x = \frac{v_0}{2}\tanh{\frac{y}{a_l}}~, \label{eq:khslvxin} \\
&& v_y = \delta v_y \exp[-(y/4a_l)^2]\sin(2\pi k_x x)~, \\
&& B_x = B_0 ~, \\
&& B_y = 0~,
\end{eqnarray}
where $v_0$ is the shear velocity, and $\delta v_y\ll v_0$ represents the amplitude of the initial perturbation, which consists of a single mode $k_x$. We also consider an ideal EoS $p=(\Gamma-1)\rho\epsilon$ with $\Gamma=5/3$.

This basic case has the advantage that there is only one initial excited mode, so that the complexity is greatly reduced and we can isolate better the effects of the SGS model implementation. The outer boundary conditions allow perturbations perpendicular to the shear layer to go out of the system, thus helping the creation of a single vortex. This corresponds to a spectrum power distributed uniquely among $k_x$ and, at much lesser extent, its higher harmonics. In other words, turbulence does not develop completely and the relevant scales to be resolved are basically set by $k_x$. Therefore, the numerical convergence is ensured for high enough resolutions $\Delta \ll 2/k_x$. Indeed, in previous works \cite{obergaulinger10,vigano19b} it is evident how the numerical resolution affects the growth rate. 

In order to evaluate convergence, we consider the evolution of the volume-integrated $y$-component of the kinetic and magnetic energies, tracers of the turbulence, and defined as in the nonrelativistic case for simplicity (relativistic corrections would not change the comparison between different resolutions):

\begin{eqnarray}\label{eq:ekiny}
 E_{ky} \equiv \int_V\frac{1}{2}\rho v_y^2~ {\rm d} V  ~~~,~~~ 
 E_{my} \equiv \int_V\frac{1}{2} B_y^2~ {\rm d} V~. 
\end{eqnarray}
After the initial triggering stage, both quantities tend to grow exponentially until a saturation level. The growth rate and the saturation value of $E_{ky}$ are physically controlled by the values of $v_0$, $a_l$, $k_x$ and $B_0$, but not by the amplitude of the perturbation, as long as $\delta v_y \ll v_0$. Analytical values of the growth rate of $E_{ky}$ are known for the nonrelativistic case only \cite{miura82}.
At saturation, the solution approaches a stationary regime, with a fully-developed vortex showing periodic oscillations in both directions, but maintaining the shape. The full development of the vortex occurs only under certain conditions of Mach number (roughly of order 1) and Alfven number (i.e., the initial large-scale magnetic field has to be weak enough to leave the fluid free to develop small structures), as explained in detail for the nonrelativistic case in~\cite{obergaulinger10}.

We set up a test with parameters $\delta v_y = 0.01$, $a_l=0.05$, $k_x=1$, $v_0 = 0.5$, and $B_0=0.0005$. With such a choice, numerical convergence of the solution is reached with only few hundreds of points in each direction, allowing fast tests (see \cite{obergaulinger10,vigano19b} for an extended exploration of parameters). In both case, we employ 3 resolutions: $N=\{25,50,100\}$. In the case of the lowest resolutions, we apply the SGS model with different values of the free parameter ${\cal C}= \{0, 1, 2, 4, 8\}$.

\begin{figure}
	\centering
	\includegraphics[width=0.45\textwidth]{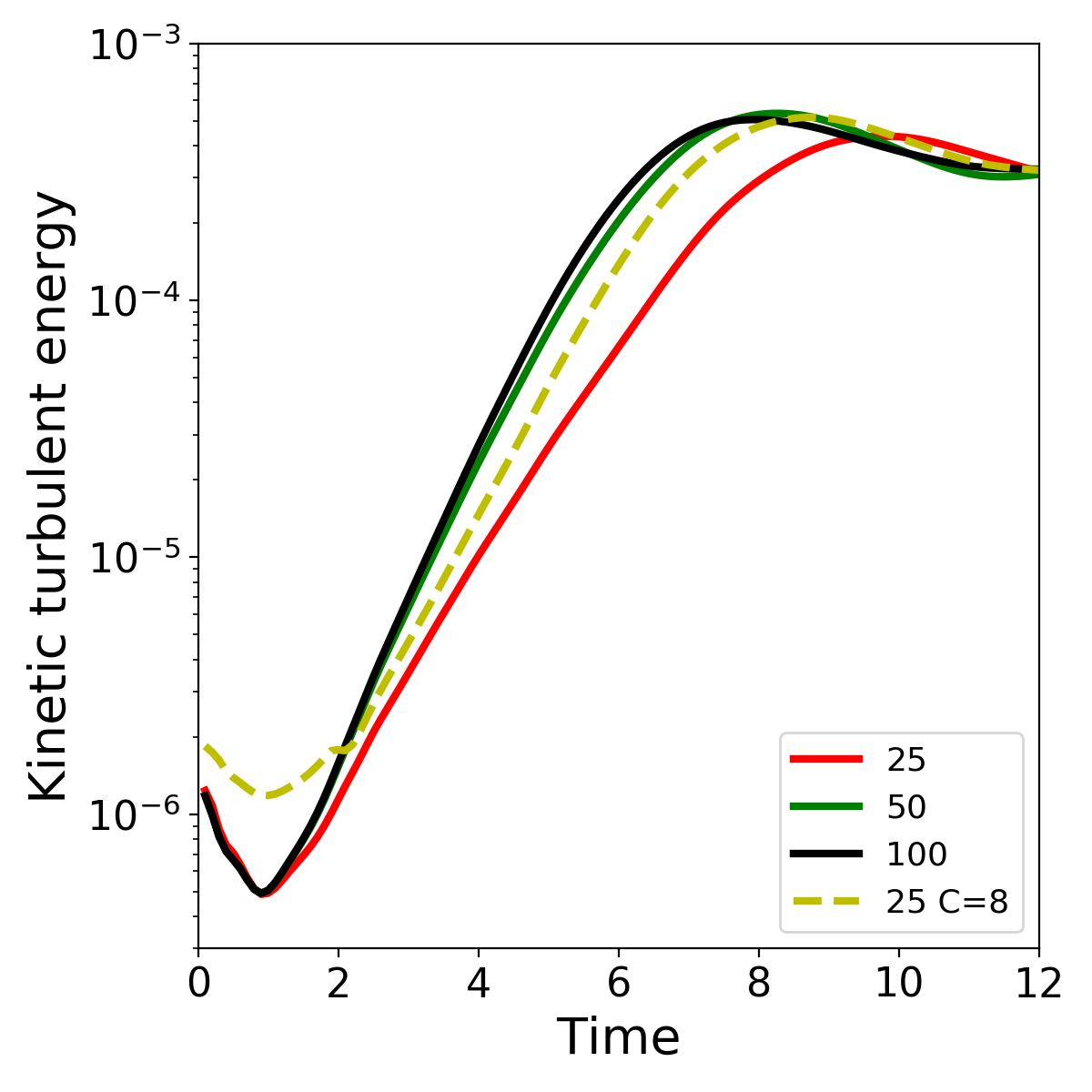} 
	\caption{{\em LES+SGS for single-vortex 2D turbulence}. Evolution of the integrated turbulent kinetic energy as a function of time. The comparison of the different resolutions without SGS model (solid lines) and the low-resolution run with ${\cal C}=8$ (dashed) indicates that the low-resolution explicit LES solution tends to the higher-resolution ones.}
	\label{fig:box_2d_singlemode}
\end{figure}

We monitor both $E_{ky}$ and $E_{my}$ as a function of time to calculate their corresponding growing rates, obtained by fitting these curves to a functional $\propto e^{\lambda_k  t}$ and $\propto e^{\lambda_m  t}$, respectively, where $\lambda$ parameters are the growth rate values. After the initial transient, the growth rate show convergence to a well-defined value as the resolution is increased (i.e., $\lambda_k = 0.97$ and $\lambda_m = 0.81$ for $N=25$, converging to $\lambda_k = 1.22$ and $\lambda_m = 1.09$ for $N=100$). Interestingly, a similar growth rate and evolution can be achieved with very low resolution $N=25$, using a sub-grid model with ${\cal C} \approx 8$ (i.e.,$\lambda_k = 1.10$ and $\lambda_m = 0.97$). In order to see some effects compared to the case without SGS, we need at least ${\cal C}=4$. The evolution of $E_{ky}$ is explicitly displayed in Fig.~\ref{fig:box_2d_singlemode} only for the extreme cases ${\cal C}= \{0, 8\}$. We also repeated the same tests in the nonrelativistic case (the models {\tt grw1-2-3-4 and 10} of \cite{obergaulinger10,vigano19b}), finding an optimal value of ${\cal C}\sim 3$. This stresses the fact that the exact value of ${\cal C}$ would need a fine-tuned calibration, which unfortunately depends on both the physical model and initial conditions and the intrinsic dissipation of the numerical methods used. However, generally speaking, the inclusion of the model (typically with ${\cal C}\sim 1-10$) will anyway improve the accuracy in reproducing the effects of the SGS dynamics.

\begin{figure*}
	\includegraphics[width=0.32\textwidth]{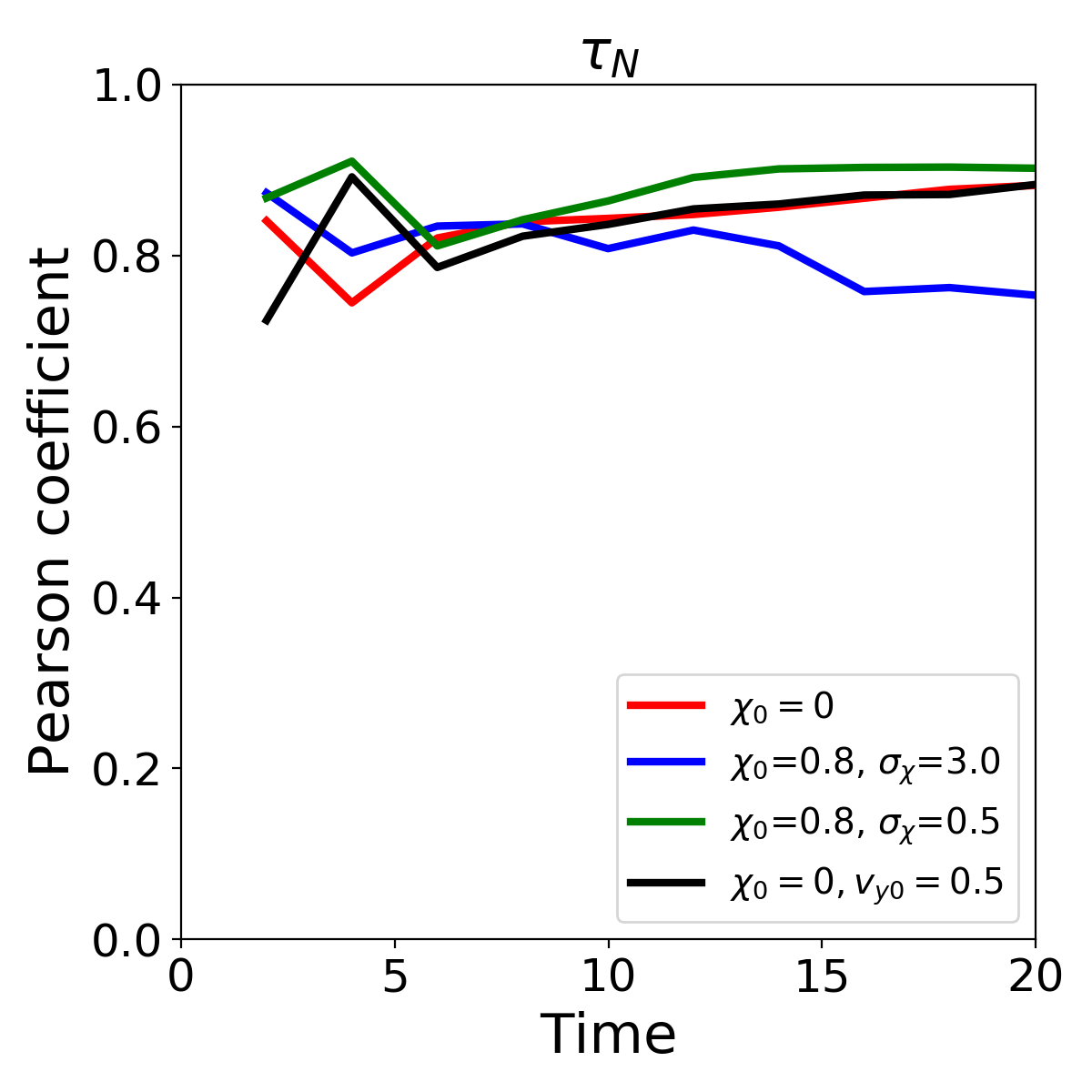}
	\includegraphics[width=0.32\textwidth]{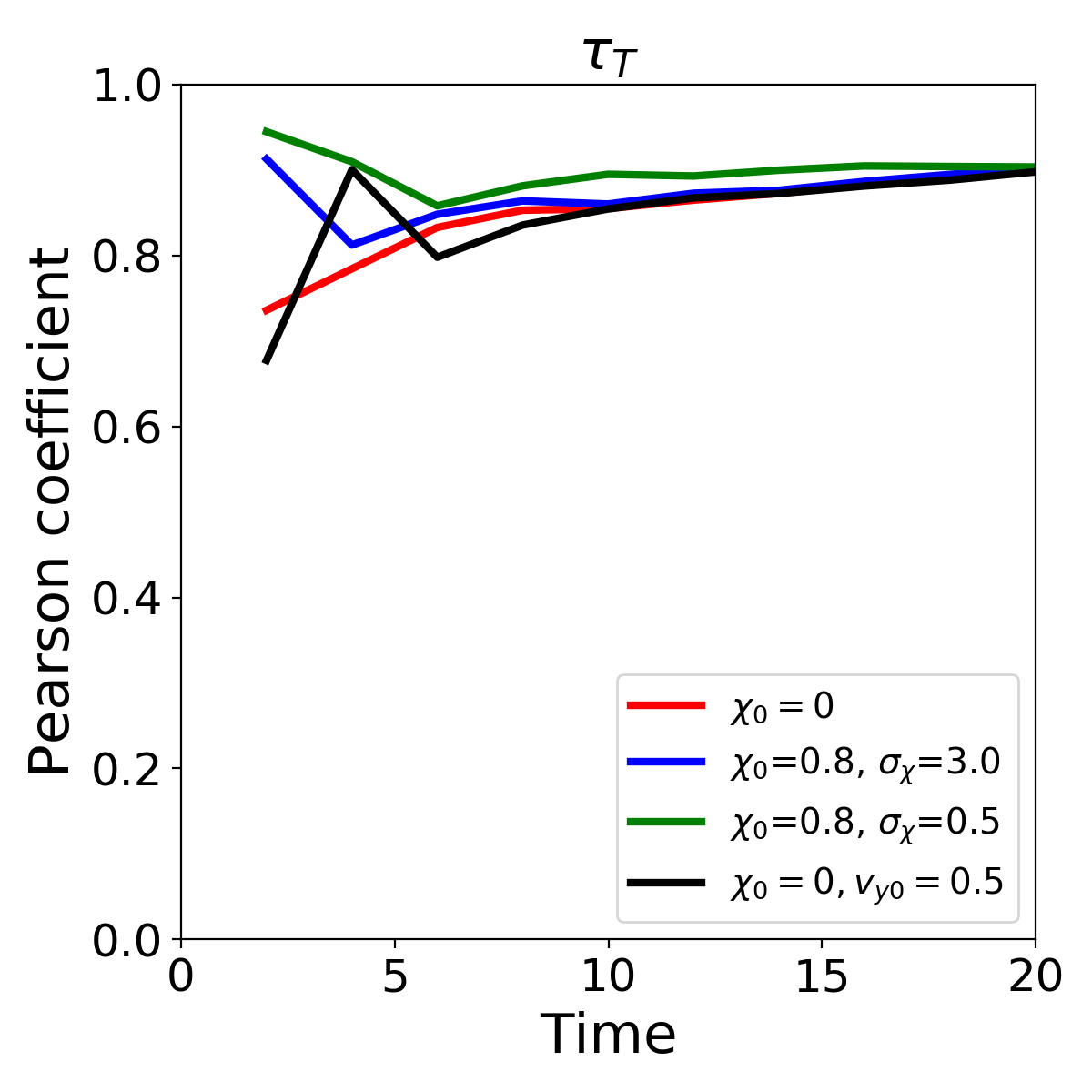}
	\includegraphics[width=0.32\textwidth]{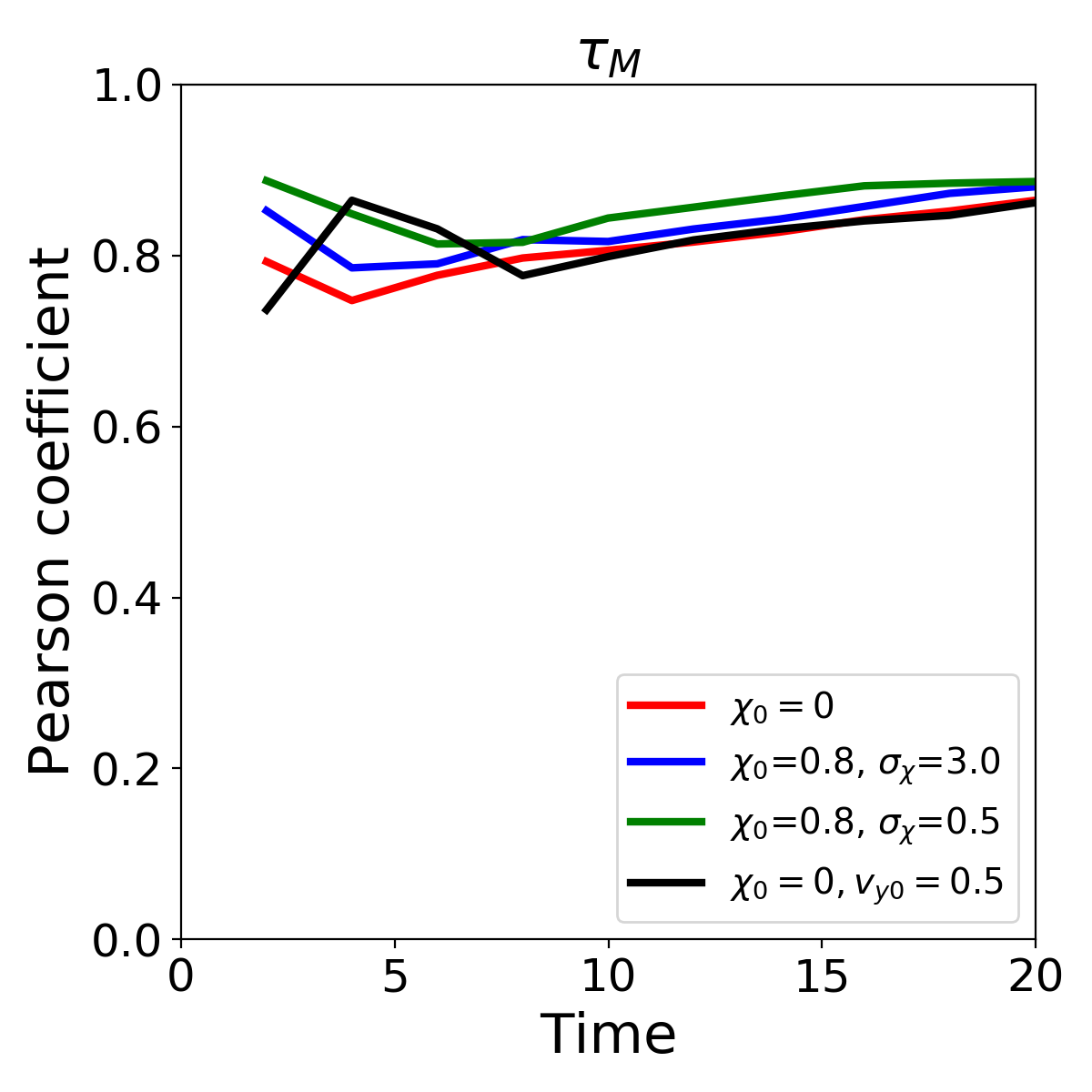} \\
	\centering
	\includegraphics[width=0.32\textwidth]{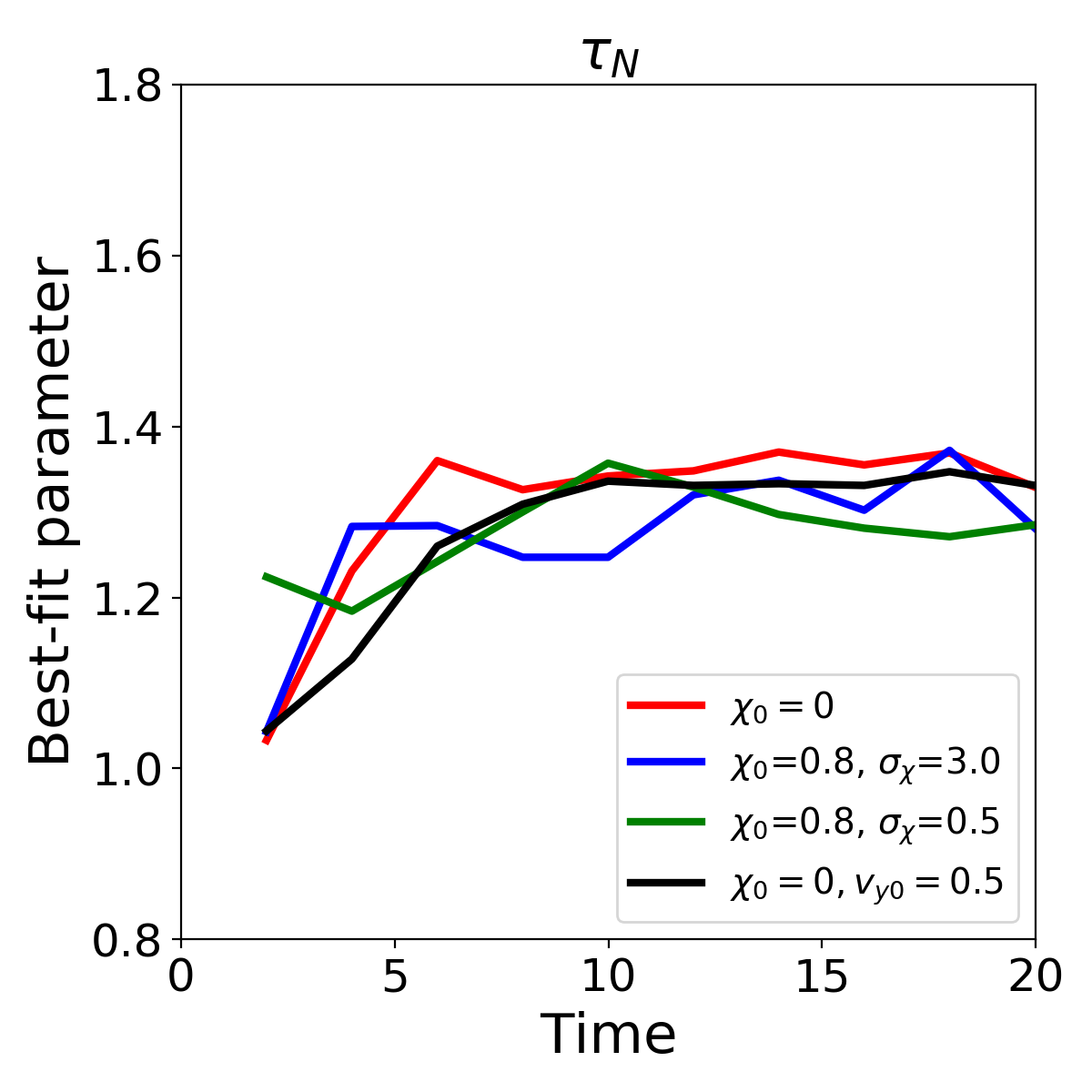}
	\includegraphics[width=0.32\textwidth]{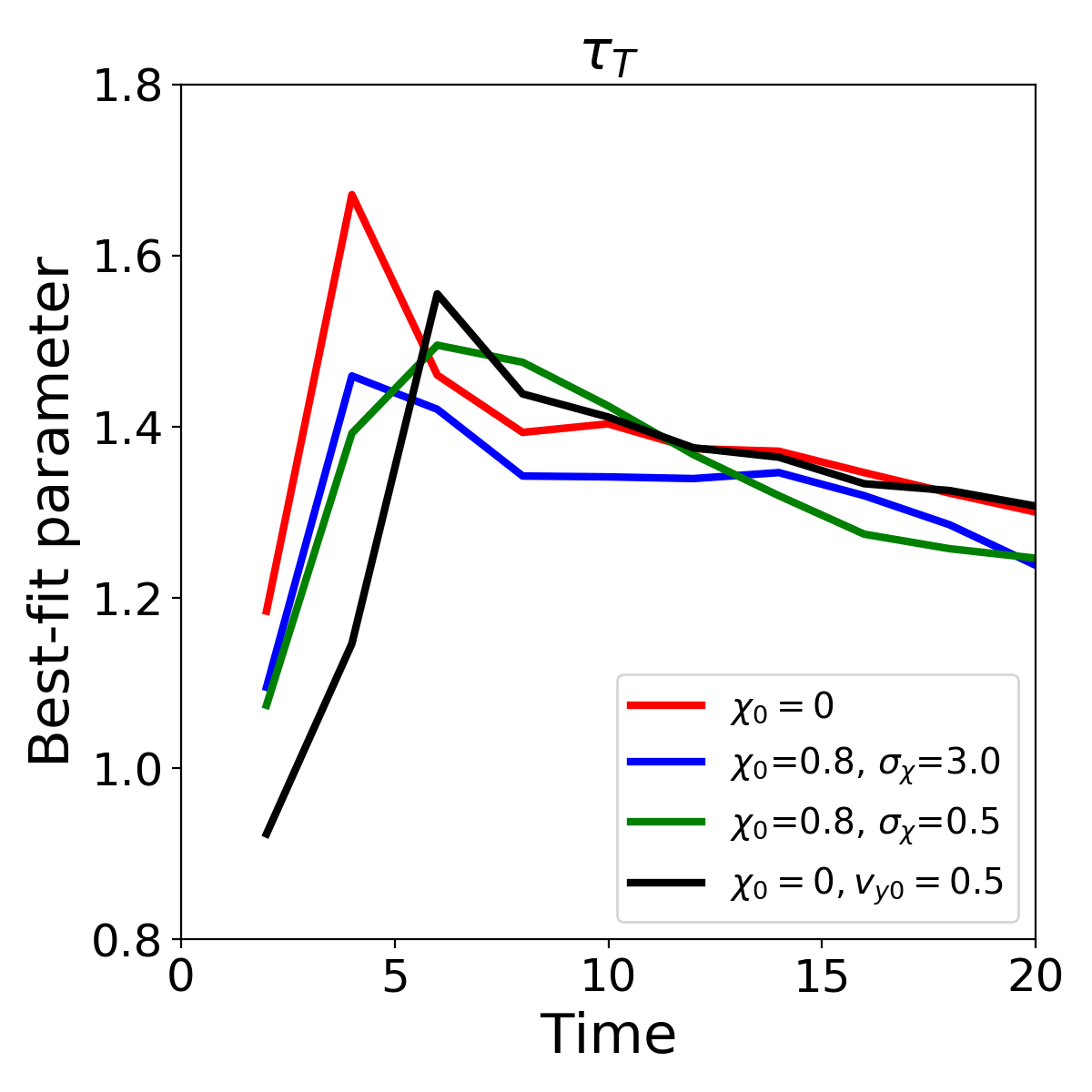}
	\includegraphics[width=0.32\textwidth]{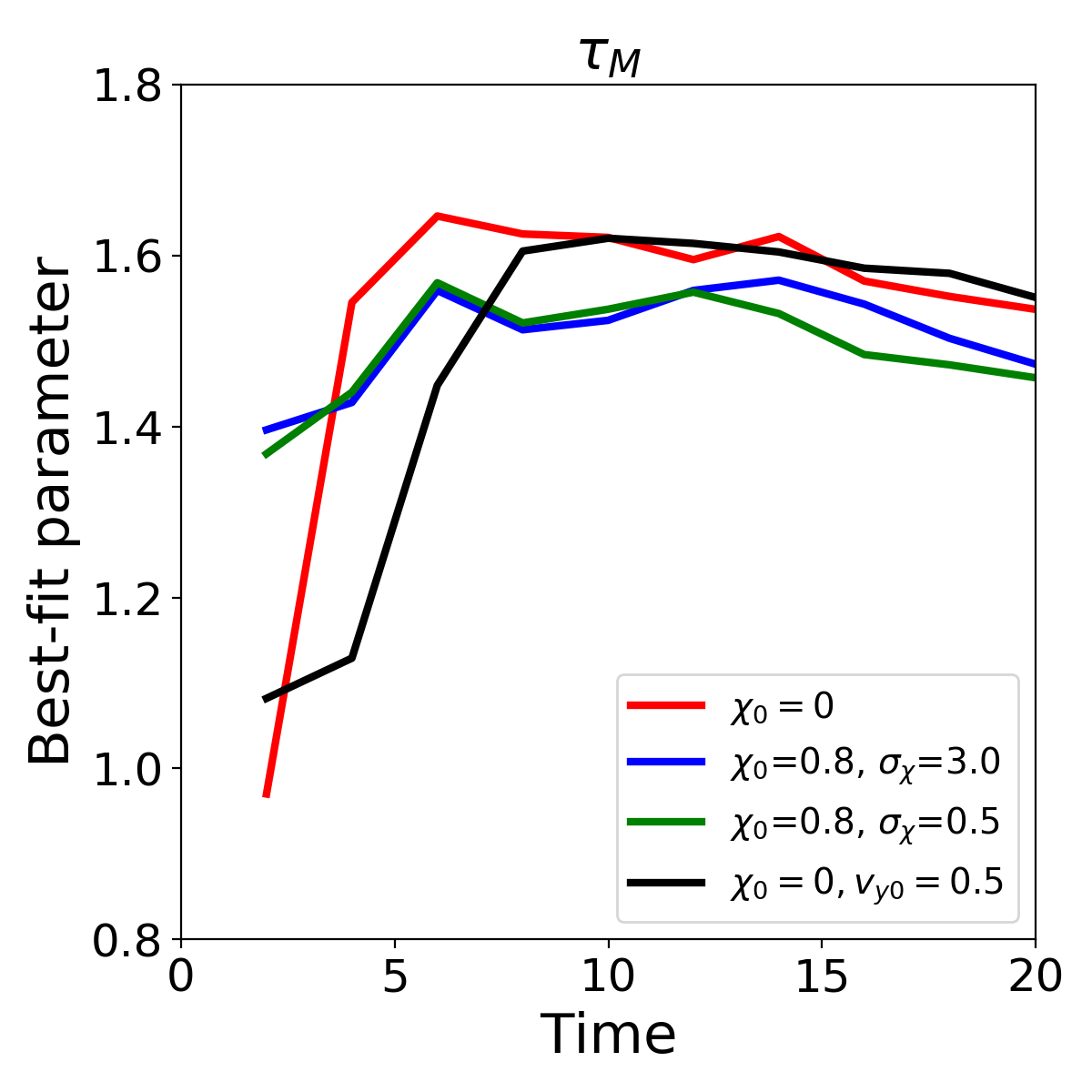}
	\caption{{\em Box simulations in 3D: a-priori tests}. Pearson coefficients (top) and best-fit values (bottom) for several cases studied with {\tt KH128C0} for a filter size $S_f=2$. Performances are not sensitive to the presence of a curved spacetime background ($\chi_0 \neq 0$) or shocks in the fluid ($v_{y0} \neq 0$).}
	\label{fig:box_pearson}
\end{figure*}

\subsection{3D fully turbulent KHI}

\subsubsection{Setup}

Next, we consider a fully turbulent 3D problem. We set our problem in Cartesian coordinates considering a periodic box $[-L/2,L/2]^3$. The primitive fields read initially:

\begin{eqnarray}
&& \rho=\rho_0 + \rho_1~\sign(y)\tanh\left(\frac{|y| - y_l}{a_l}\right)~, \\
&& v_x = v_{x0} ~\sign(y)\tanh\left(\frac{|y| - y_l}{a_l}\right) + \delta v_x ~,\\
&& v_y = v_{y0} ~\sign(y)\tanh\left(\frac{|y| - y_l}{a_l}\right)\nonumber\\
& &~~~~~~+ \delta v_y~\sign(y)  \exp\left[-\frac{(|y| - y_l)^2}{\sigma_y^2}\right] ~,\\
&& v_z = v_{z0}~\sign(y)\exp\left[-\frac{(|y| - y_l)^2}{\sigma_z^2}\right] + \delta v_z ~,\\
&& \vec{B} = B_{x0} \hat{x} ~,~
 p=p_0 ~,
\end{eqnarray}
where $a_l$ is the mixing layer scale, $y_l$ is the distance of the shear layers to the plane $y=0$, $\sigma_y$ and $\sigma_z$ are the extension scale of the initial perturbation in the $y$-direction and the profile of $v_z$, respectively. The main flow is initially given by $v_{x0}$ and $v_{y0}$, having opposite signs across the layer. The standard values that we consider are $L=1$, $y_l=1/4$, $a_l=0.01$, $v_{x0}=0.5$, $v_{y0}=0$, $p_0=1$, $B_{x0}=0.001$ and $\sigma_z^2=0.1$. The initial perturbation, $\delta v_i$, is a superposition of single-mode perturbation with a number of nodes $n_i \in [1,N/2]$, periodic in the boundary box, $\delta v_i=\delta v_0 \sin(2\pi x_i n_i/L)$, with $n_x=11$, $n_y=7$, $n_z=5$ (different prime numbers-based initial modes in each direction ensure the excitation of the entire spectrum of modes, see \cite{vigano19b} for a discussion), $\delta v_x = \delta v_z = 0.01$, $\delta v_y = 0.1$, $\sigma_y^2=0.01$. We employ an ideal gas EoS with $\Gamma = 4/3$.

\begin{figure*}
	\includegraphics[width=0.32\textwidth]{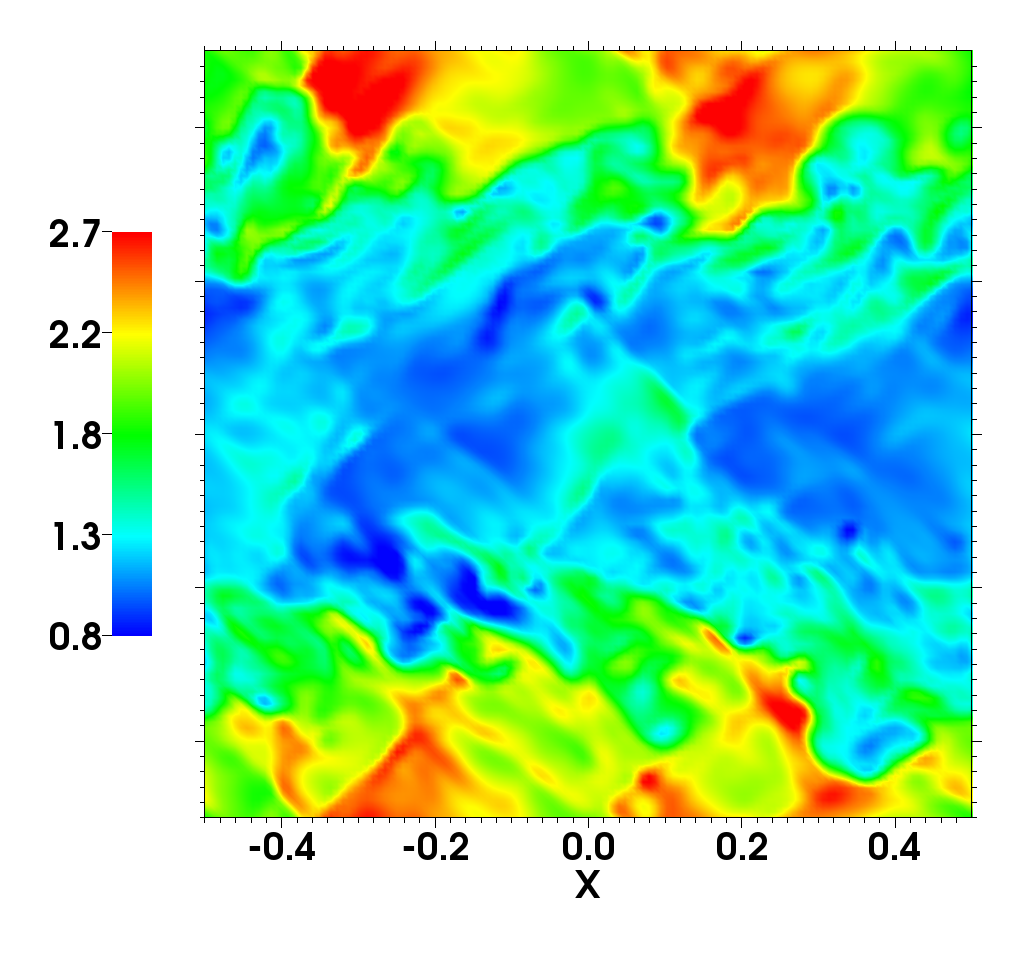}
	\includegraphics[width=0.32\textwidth]{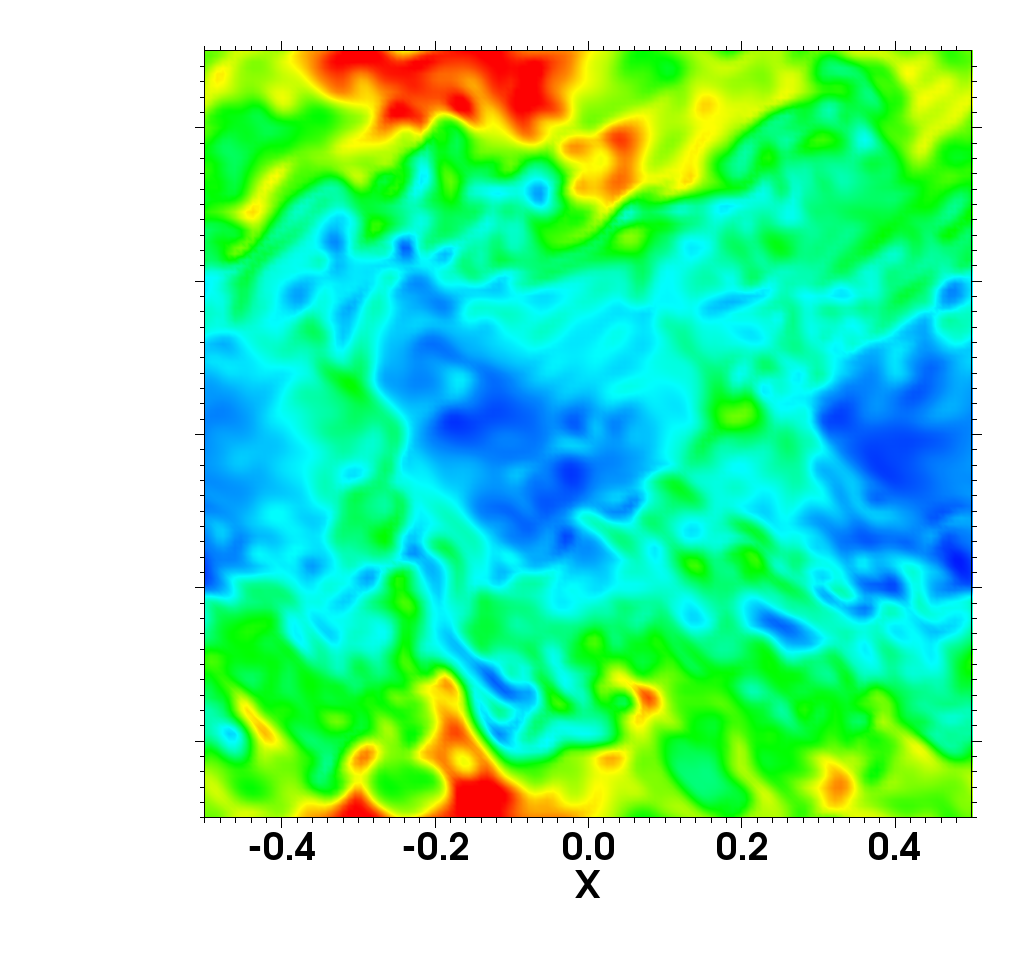}
	\includegraphics[width=0.32\textwidth]{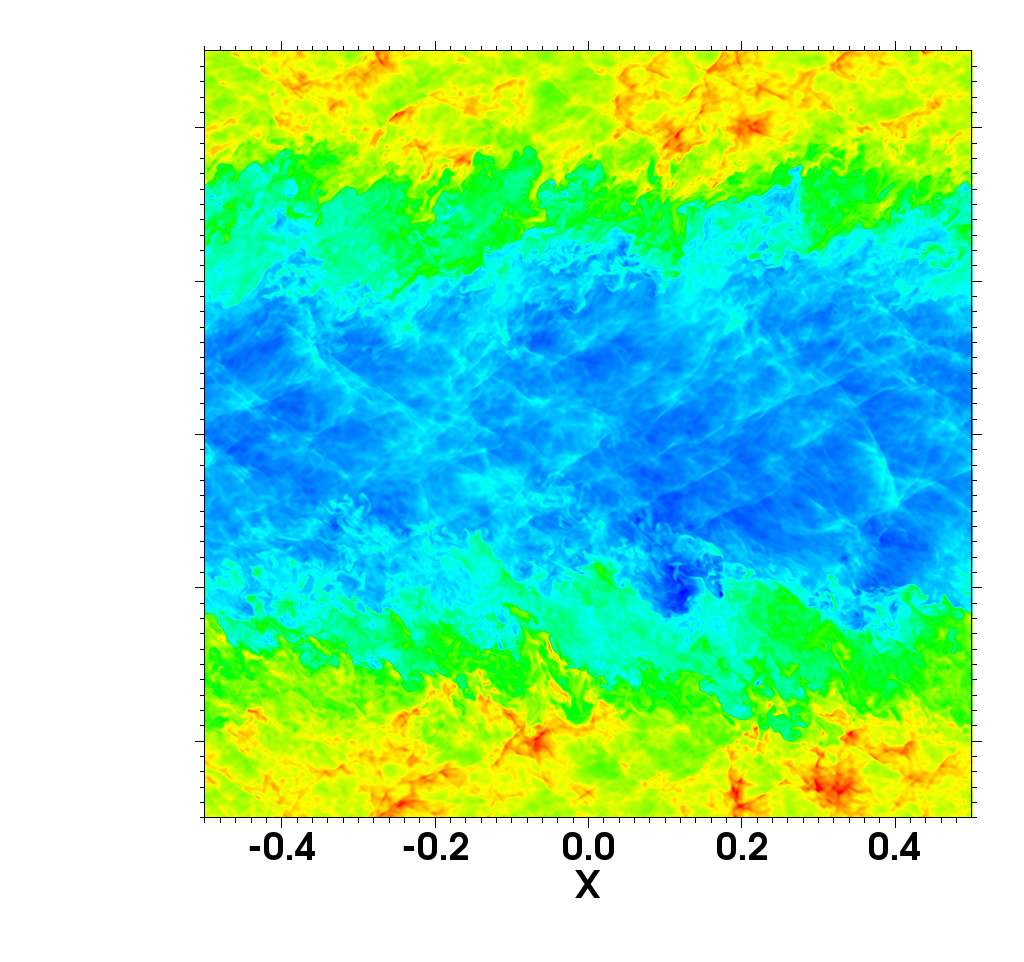} \\
	\centering
	\includegraphics[width=0.32\textwidth]{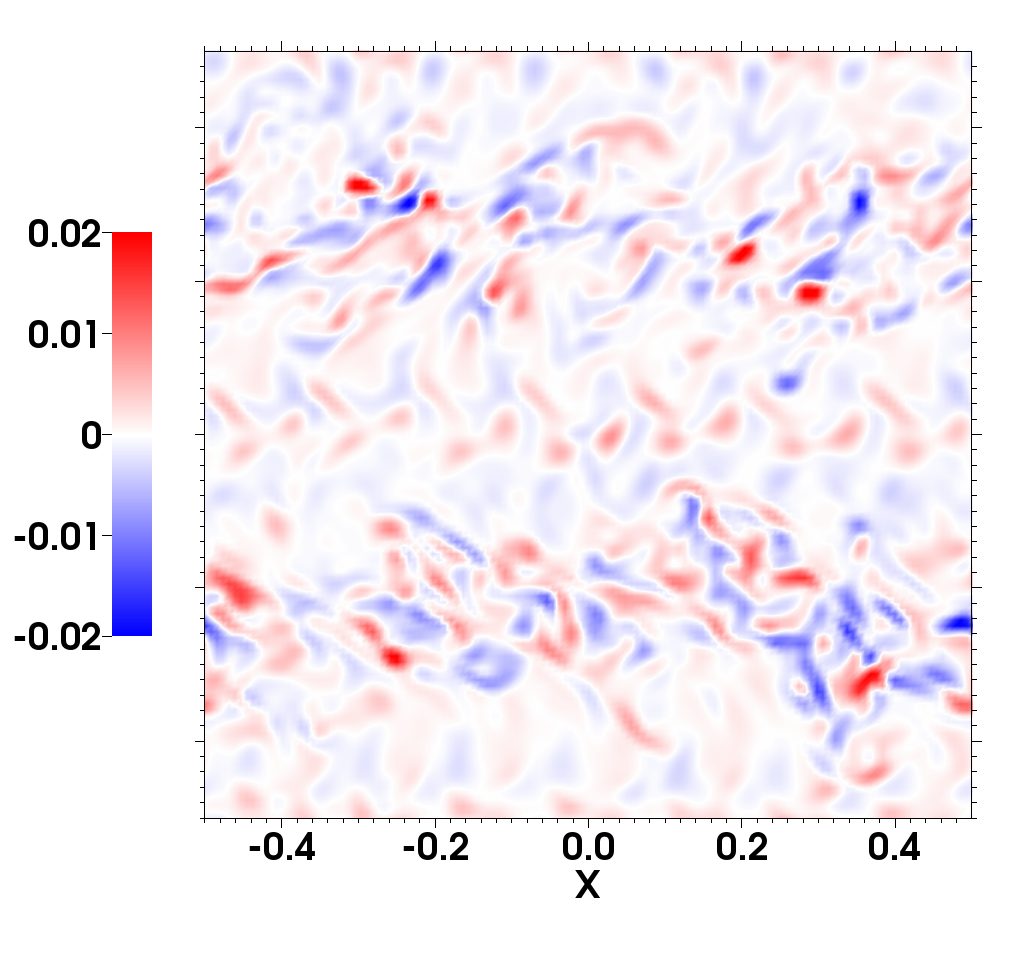}
	\includegraphics[width=0.32\textwidth]{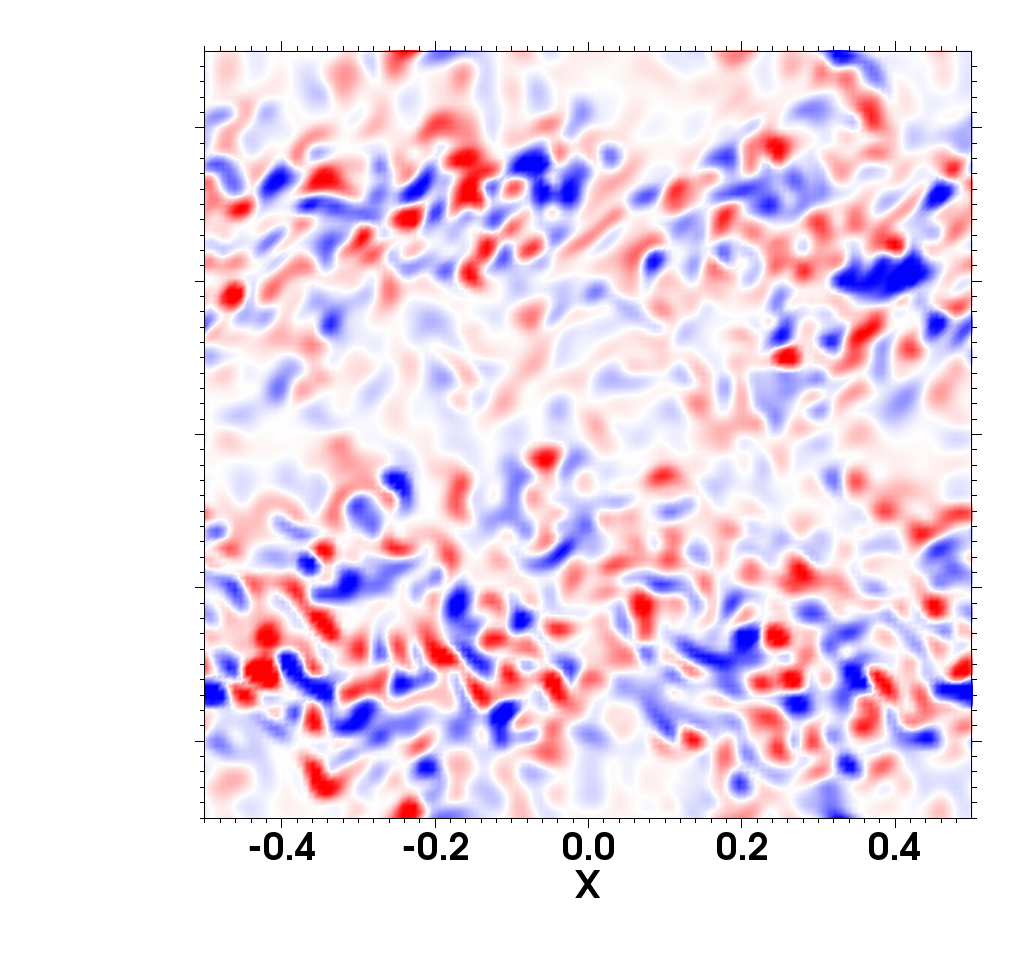}
	\includegraphics[width=0.32\textwidth]{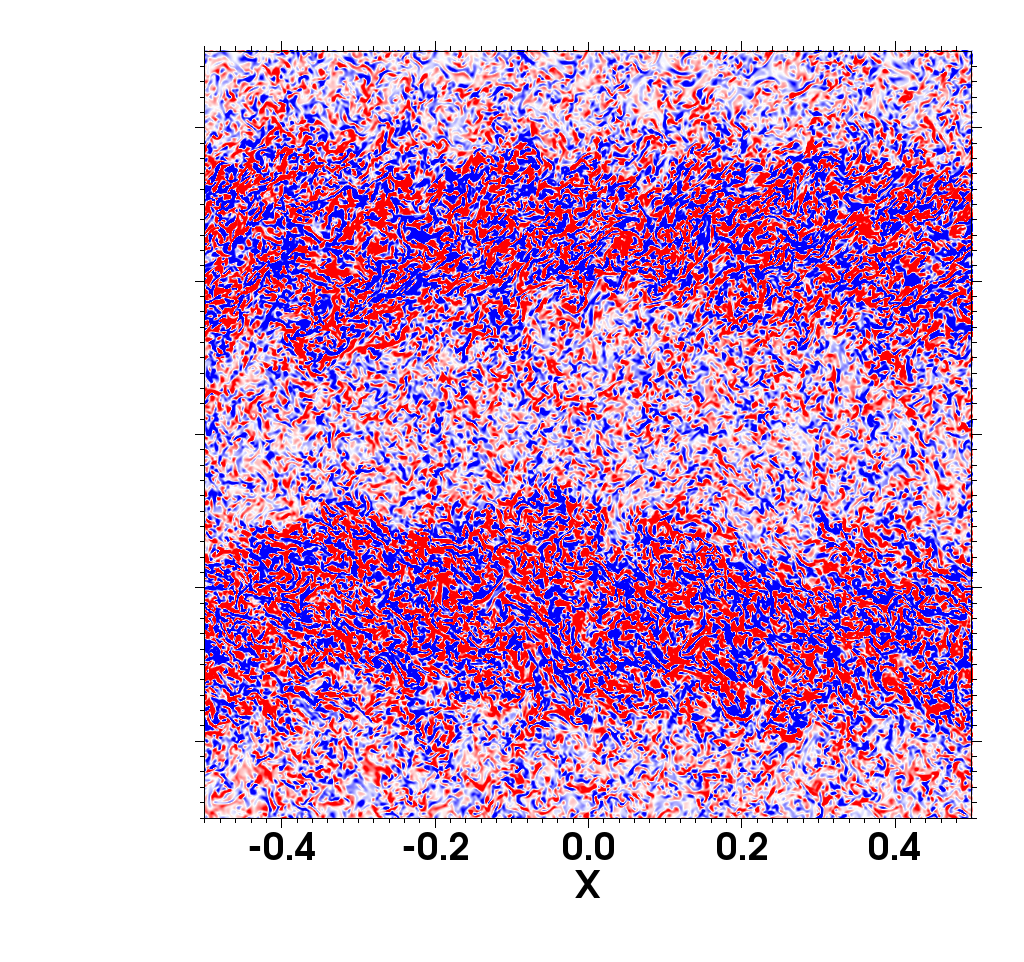}
	\caption{{\em Box simulations in 3D}. Snapshots of the z-component of the magnetic field, $B^z$,  at $t=6$ for $N=128$ and ${\cal C}=0$ (left), $N=128$ and ${\cal C}=8$ (center) and $N=1024$ and ${\cal C}=0$ (right). Turbulence has not completely developed at this time, but eddies have begun to form, especially at the highest resolution.}
	\label{fig:box_densiy_Bz}	
\end{figure*}

As noted in previous works, the fully-developed (i.e., coming from a random/multimode initial perturbation) the KHI is known to grow faster for smaller scales, and due to the absence of physical viscosity in this benchmark test, there is no physical lower limit to the scale of the excitable modes. Since our initial perturbation is designed to excite the entire spectrum of modes, we do not expect a numerical convergence in the growth phase, since the more we refine the grid, the more fast-growing excited modes will be included.

\begin{figure}
	\centering
	\includegraphics[width=0.42\textwidth]{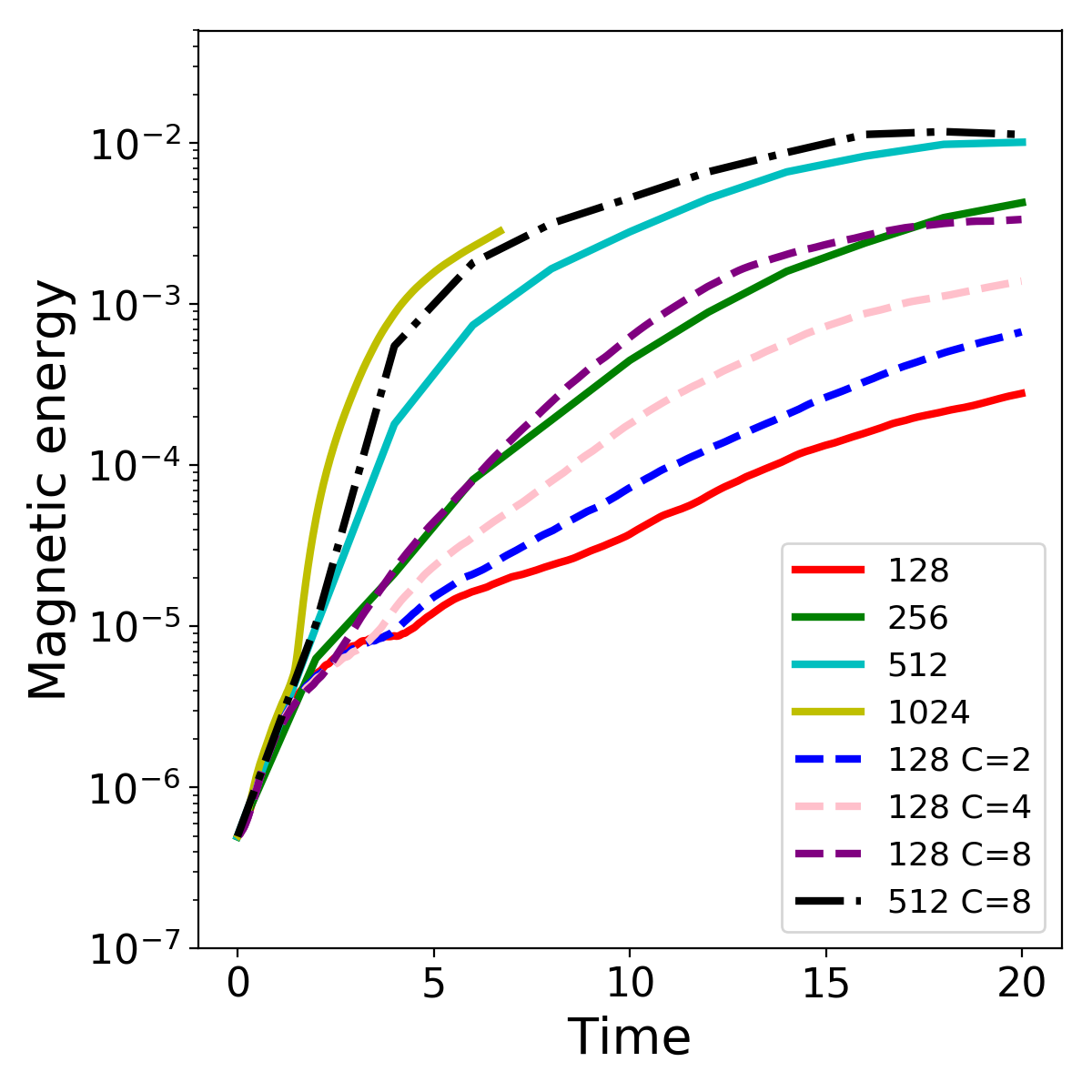}
	\caption{{\em Dynamo in 3D KHI}. Total magnetic energy as a function of time for different resolutions and dependence on the free parameter ${\cal C}$ of a box simulation. The magnetic energy rises as we increase ${\cal C}$ and/or the resolution.}
	\label{fig:box_emag}
\end{figure}

\begin{figure*}
	\centering
	\includegraphics[width=0.246\textwidth]{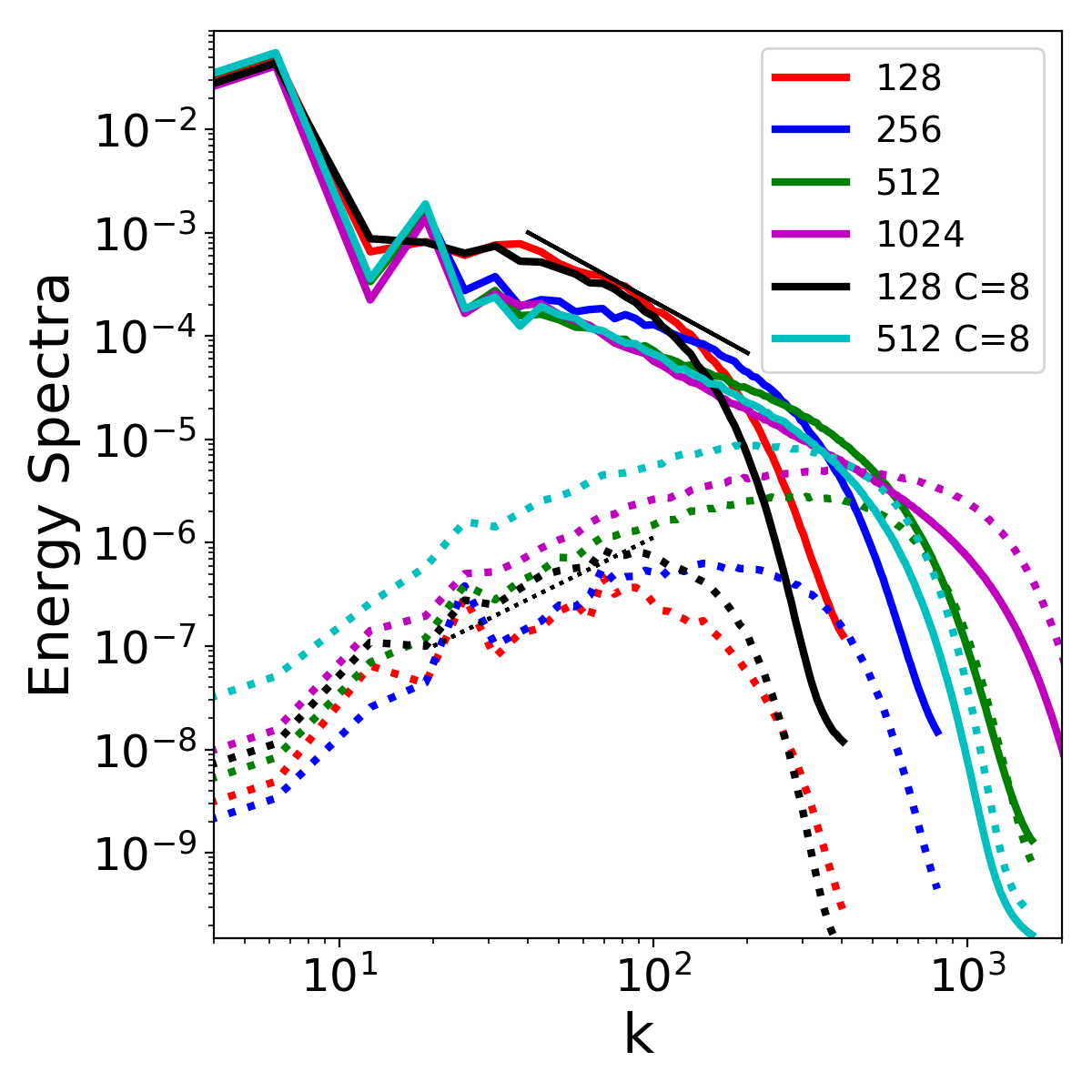}
	\includegraphics[width=0.246\textwidth]{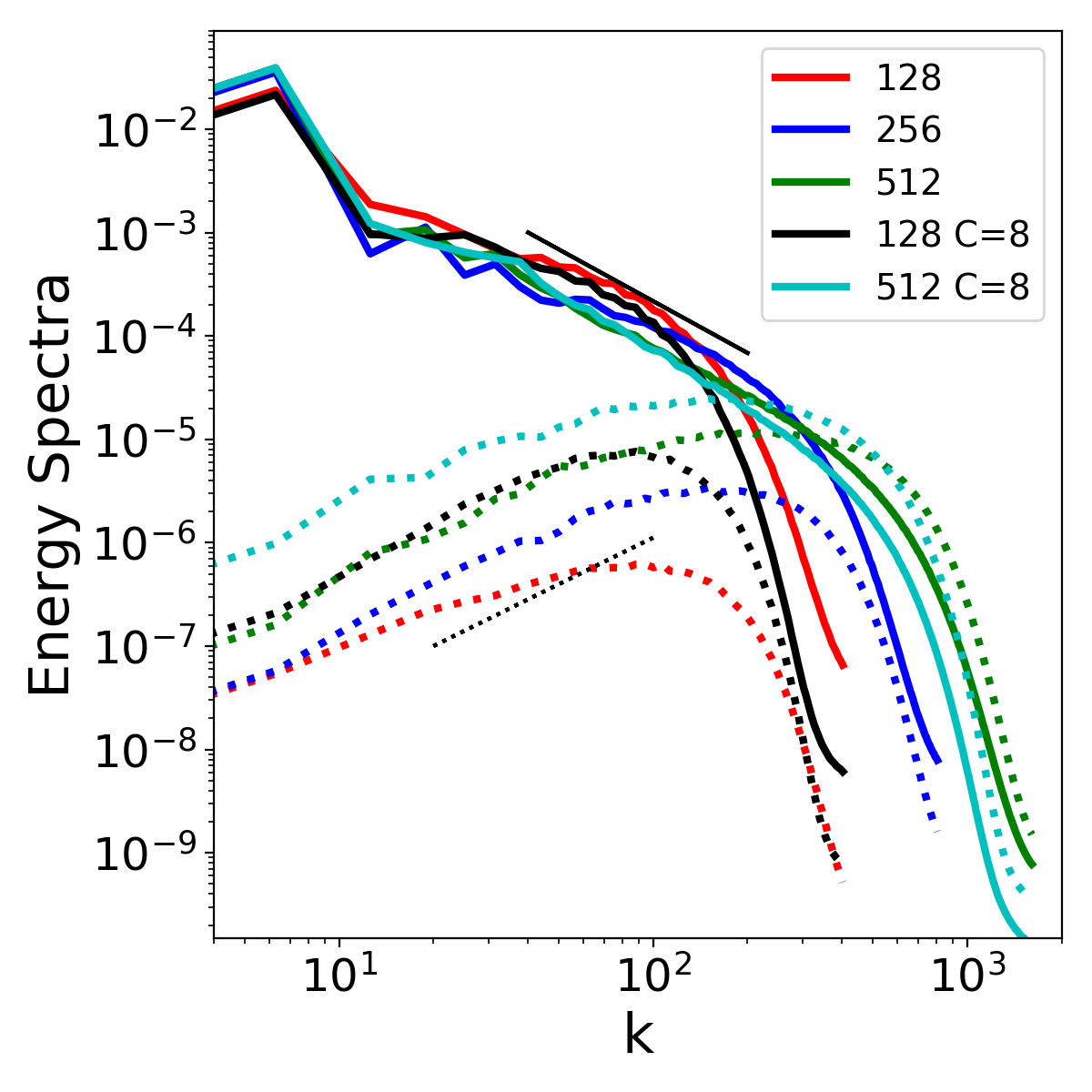}
	\includegraphics[width=0.246\textwidth]{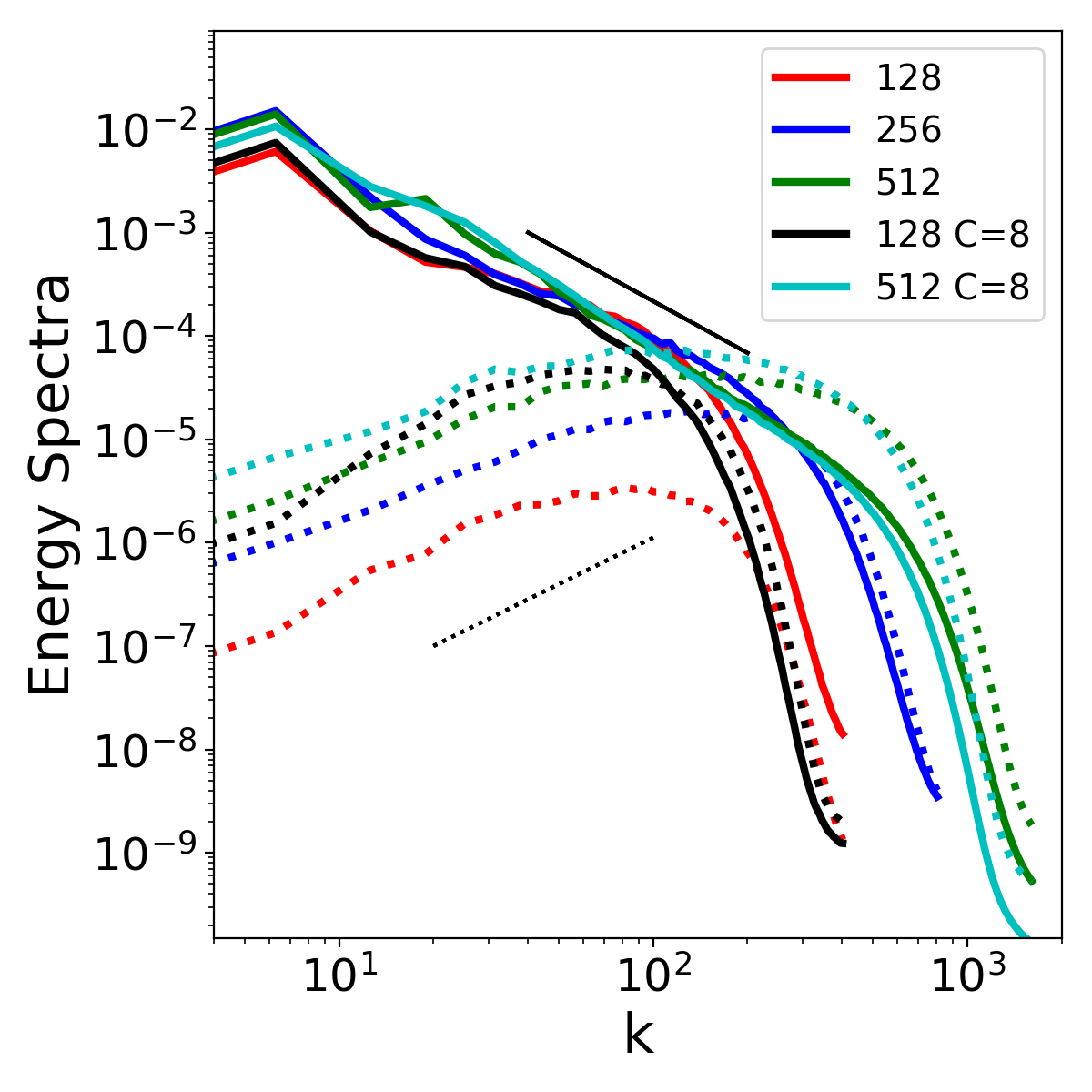}
	\includegraphics[width=0.246\textwidth]{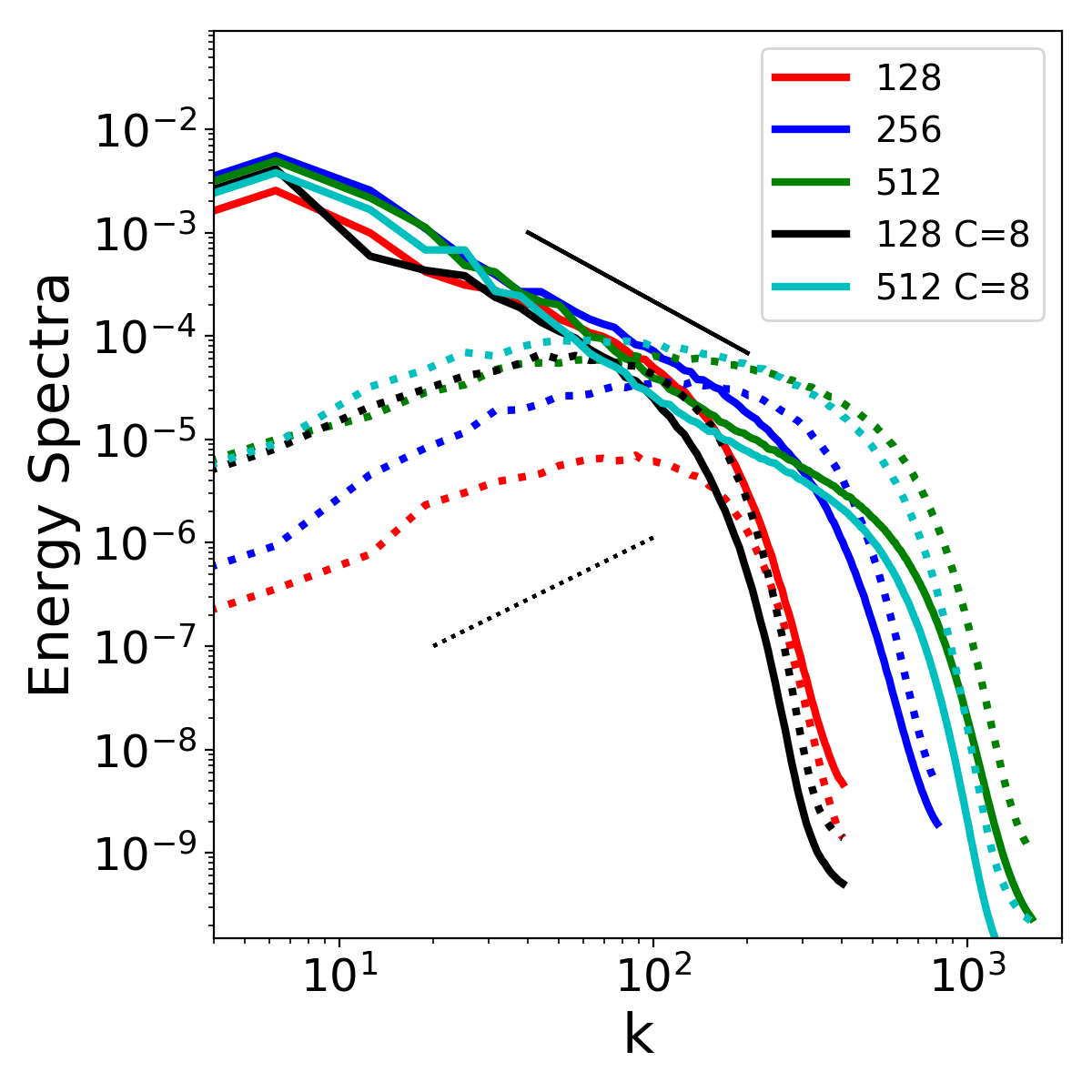} 	
	\caption{{\em Box simulations in 3D: a-posteriori tests in Minkowski spacetime}. Kinetic (solid) and magnetic (dashed) energy spectra at time-slices (from left to right) $t=\{6,10,16,20\}$ of a box simulation, for different models without or with SGS models (${\cal C}=8$). The black, thin solid line and dots represent the Kolmogorov $\propto k^{-5/3}$ and Kazantsev $\propto k^{3/2}$ slopes, respectively, as references.}
	\label{fig:box_spectra}
\end{figure*}

We validate the gradient SGS model for GRMHD under a broad range of physical conditions, extending some of the tests already performed in previous works \cite{vigano19b,carrasco19}. First, we vary different values of the maximum density $\rho_0 + \rho_1 \in [2:50]$ (keeping the minimum density $\rho_0 - \rho_1 = 1$), representative for small to extreme initial density jumps. Secondly, we consider a fixed, non-flat metric by
setting the conformal factor to a Gaussian shape,
\begin{equation}
	\chi= 1 - \chi_0\, e^{-(r/\sigma_\chi)^2}~,
\end{equation}
with given depth $\chi_0$ and width $\sigma_\chi$, being $r$ is the radial distance to the center of the box. We consider depth up to $\chi_0=0.8$, which is roughly the value reached in realistic BNS merger simulations. We set the width either to $\sigma_\chi=0.5$ or $\sigma_\chi=3.0$, leading to a total variation of $\chi$ between $\chi=[0.20-0.90]$ and $\chi=[0.20-0.24]$, respectively.
Thirdly, we test different values of the initial vertical velocity $v_{y0}$, such that strong shocks, traveling along the y-direction, are initially induced in the fluid.

\subsubsection{A-priori tests}

The results for the {\texttt a-priori} tests are summarized in Fig. \ref{fig:box_pearson}, displaying ${\cal P}$ (top) and the best-fit ${\cal C}$ (bottom). In all these cases, for any given time the \textit{a-priori} tests give ${\cal C}\sim 1-2$ and ${\cal P}\gtrsim 0.8$ for the minimum filter factor $S_f=2$. Performances gradually degrade for increasing loss of information, e.g. higher $S_f$ (see \cite{vigano19b,carrasco19} for a detailed exploration of other values of $S_f$ and $N$).

This first test further confirms that the gradient SGS model is still valid when a high density jump is included, when shocks are present on the solution, or when non-trivial metric factors are incorporated. More importantly, it validates the smooth-metric assumption and reinforces that the presence of a curved spacetime background does not compromise the good performance of the model. 

\subsubsection{A-posteriori tests}

We analyze now the simulations performed by including the
gradient SGS model in the evolution equations, that is, with ${\cal C} \geq 1$ (if not specified, the same indicated value of ${\cal C}$ is applied to all SGS terms). For concreteness, we  focus on the case with a flat metric background and $\rho_0=1.5$, $\rho_1=0.5$. We perform simulations with no SGS modeling (${\cal C}=0$) for different resolutions $N=\{128, 256, 512, 1024\}$, and four LES+SGS with a low resolution, $N=128$, for ${\cal C}=\{1,2,4,8\}$. We also perform a high-resolution simulation $N=512$ with ${\cal C}=8$ to check the convergence of solutions when LES and SGS modeling are included. Note that higher values of ${\cal C}$ result in numerical instabilities, besides arguably becoming physically and mathematically not consistent.
From now on, we will label the different simulations through the values of $N$ and ${\cal C}$, e.g. {\tt KH128C4} for $N=128$ and ${\cal C}=4$. 

In Fig.~\ref{fig:box_densiy_Bz} we show the density distribution and the z-component of magnetic field $B^z$, which develops due to the turbulent instability, at $t=6$ for {\tt KH128C0}, {\tt KH128C8} and {\tt KH1024C0}. As expected, the intensity of the magnetic field rises as the value of ${\cal C}$ is increased (i.e, compare the left and middle panels). The same effect is achieved by increasing the resolution (right panel), getting up to an order of magnitude higher intensities than in the lower resolution. Although the global effects are similar, notice that the {\tt KH128C8} case cannot show fine details due to its intrinsic low resolution: this does not prevent it from reproducing the feedback unresolved dynamics, thanks to the SGS terms.

In Fig.~\ref{fig:box_emag} we can see this trend quantitatively: the  volume-integrated magnetic energy rises more if we increase either the resolution or the value of ${\cal C}$. A quasi-stationary state, indicated by saturation of the integrated magnetic energy, is only achieved at the end of the simulation in the high-resolution cases and in the {\tt KH128C8}. This means the turbulence has been completely developed at this time. Nota that in the {\tt KH128C8} case we get a magnetic energy roughly an order of magnitude above the {\tt KH128C0} one and it is similar to {\tt KH256C0}. This means we are capturing the same magnetic energy with half of the resolution. The magnetic energy difference between the {\tt KH512C0} and {\tt KH512C8} simulations is less than the difference between the {\tt KH128C0} and {\tt KH128C8} ones, showing two important results: (i) the SGS effects actually converge to zero as the resolution increases, and (ii) the solutions of the simulations with the SGS modeling actually converge to those obtained with higher resolutions, but using much less computational resources.

In Fig.~\ref{fig:box_spectra} we show the spectra of both magnetic (dashed line) and kinetic (solid line) energy at different time snapshots corresponding to $t=\{6,10,16,20\}$. To improve the visualization of the results, we do not include the ${\cal C}=\{1,2,4\}$ cases with $N=128$: they lie in the middle between ${\cal C}=0$ and 8. The maximum of the energy spectra is reached before the spectral knee (located at $k$ a few times smaller than $k_{\rm max}\equiv \pi/\Delta$), after which the slope steepens due to the intrinsic dissipation of the finite-difference method. At $t=6$ we are still at the beginning of the simulation and eddies have not been developed completely yet. For this reason, almost all magnetic energy spectra have similar values. At $t=10$ we can clearly see the effects of the SGS model in the {\tt KH128C8} simulation, whose magnetic energy spectra arises above the {\tt KH256C0} case and reaches similar values to the {\tt KH512C0} one. At $t=16$, the spectra of the high-resolution case {\tt KH512C0} and the {\tt KH128C8} are getting closer at large scales (low wave-numbers). At $t=20$ the system has reached a quasi-stationary state and the energy spectra is comparable for the highest resolution simulations {\tt KH512C0}, {\tt KH512C8} and the lowest resolution one with SGS modeling {\tt KH128C8}. Notice that at all times the spectra of {\tt KH512C8} has still the highest magnetic energy, as it combines the small scale dynamics (i.e., accessible because of the high resolution) and the SGS effects.

\begin{figure*}
	\includegraphics[width=0.32\textwidth]{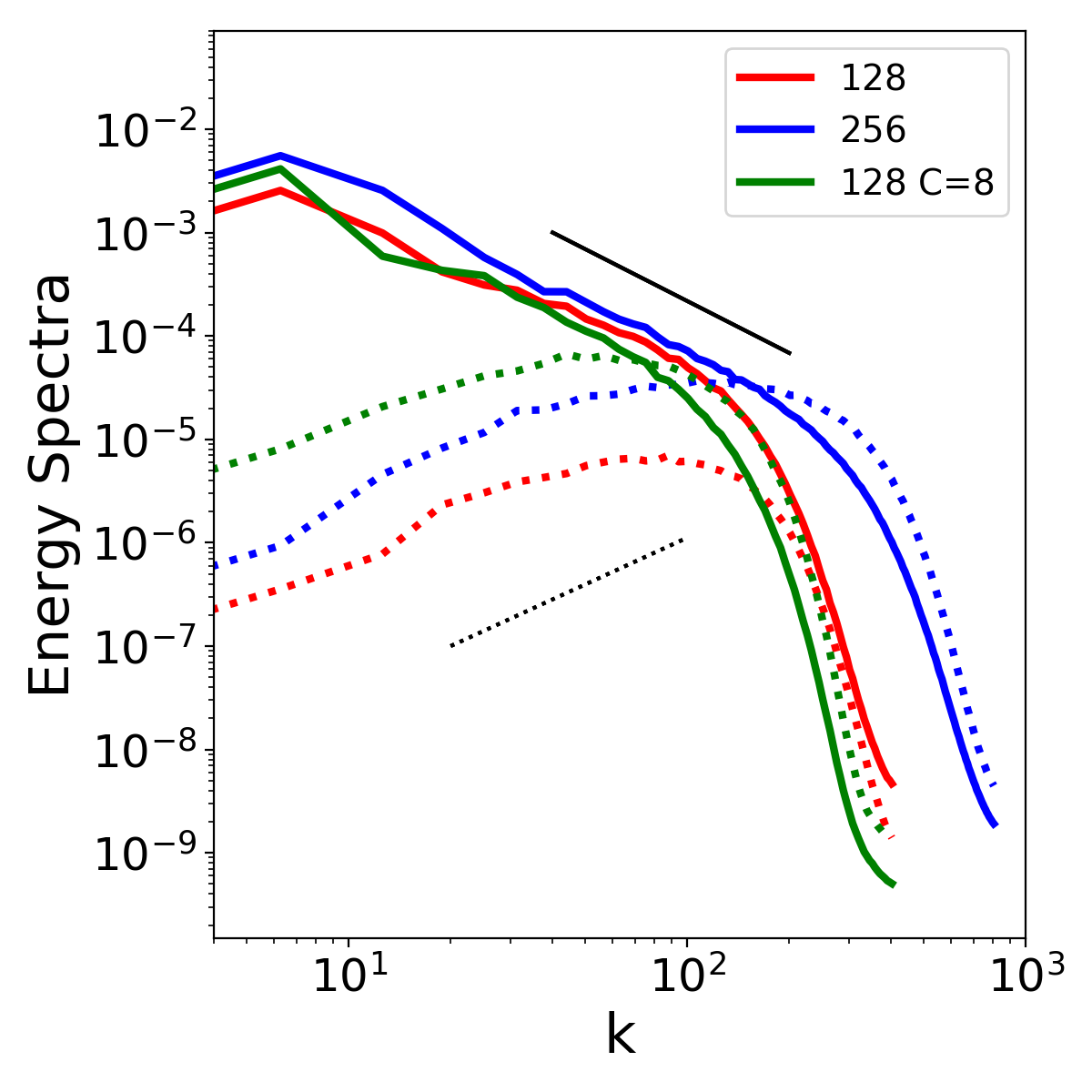} 
	\includegraphics[width=0.32\textwidth]{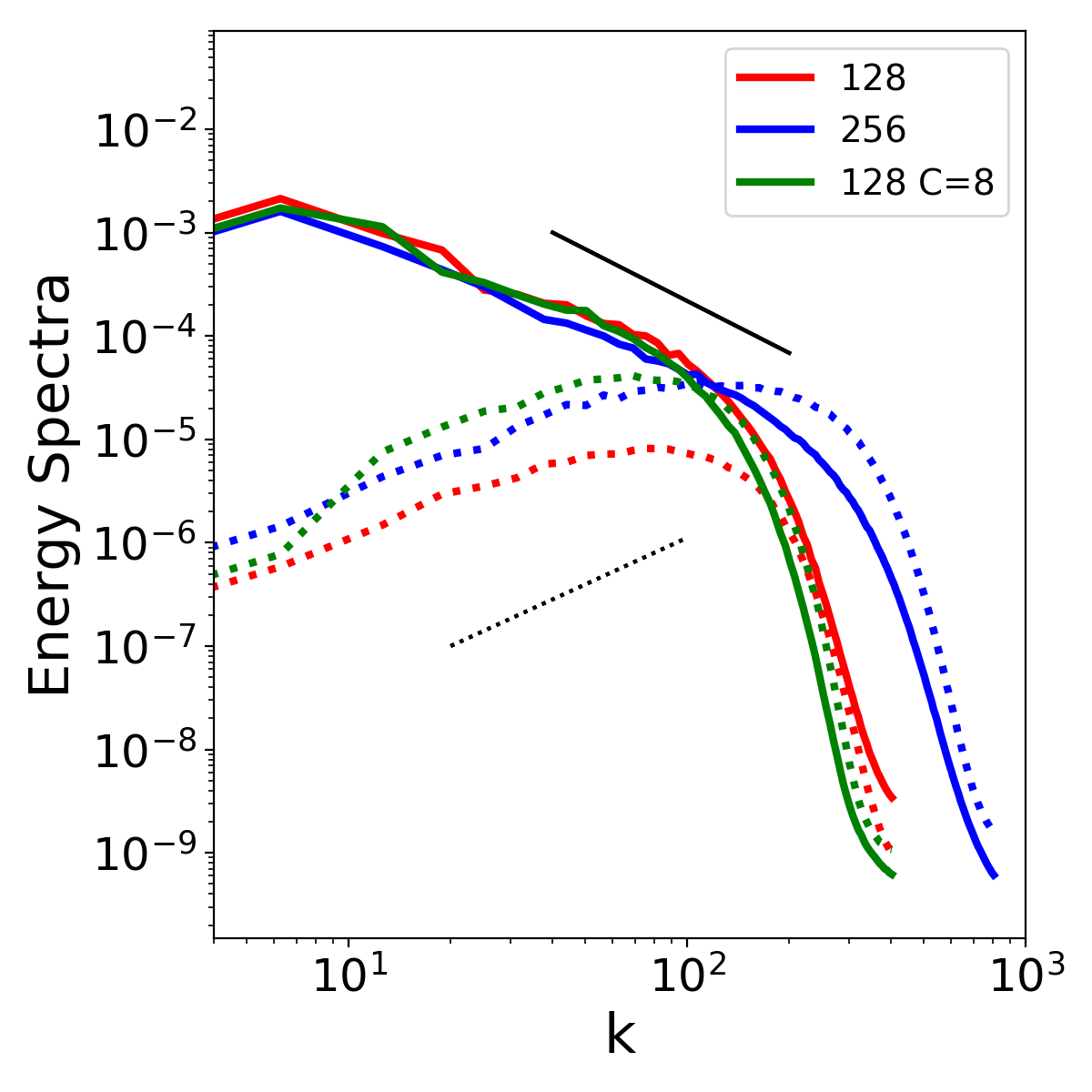} 
	\includegraphics[width=0.32\textwidth]{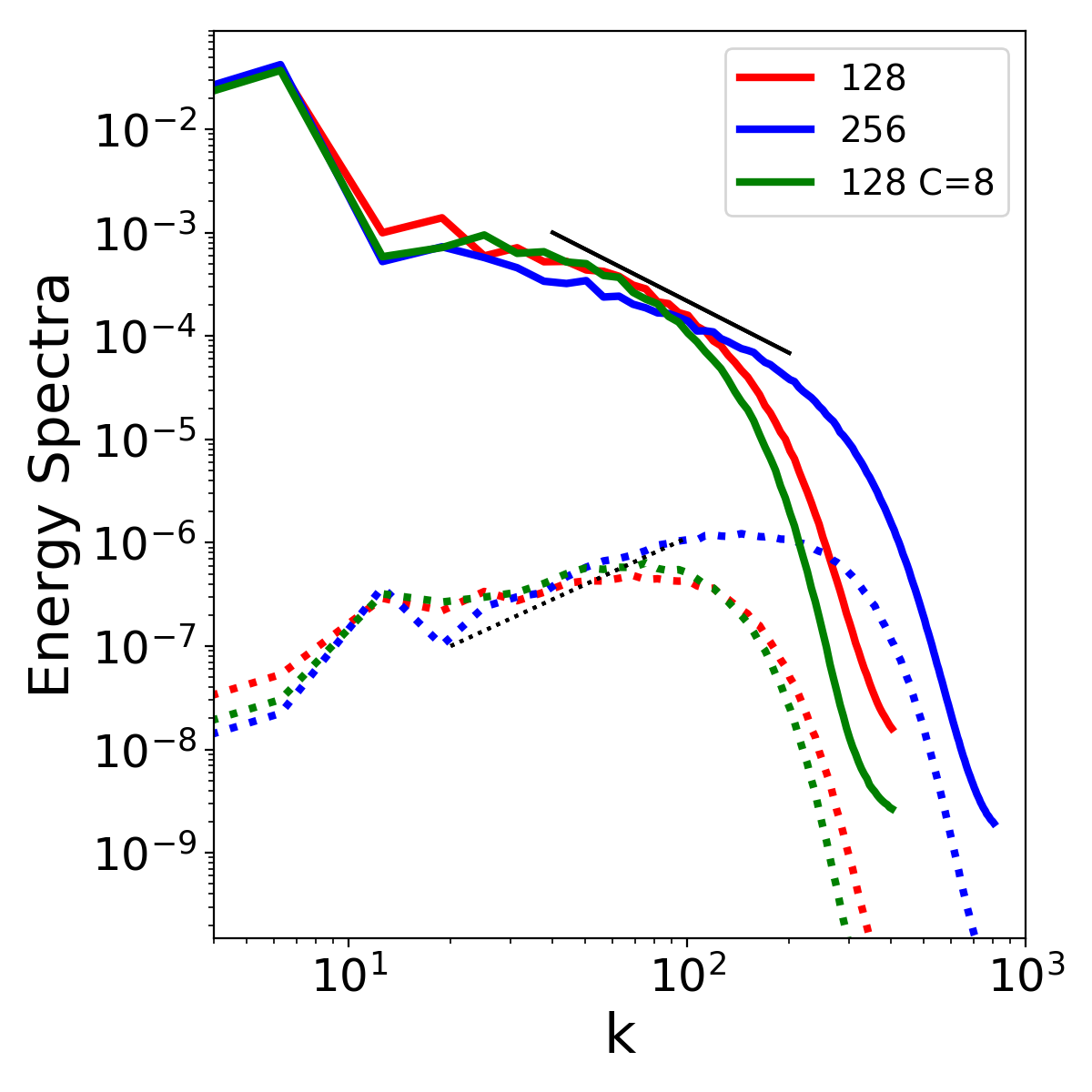} 	
	\caption{{\em Box simulations in 3D: a-posteriori tests in curved background}. Kinetic (solid) and magnetic (dashed) energy spectra at $t=20$ for $\chi=1$ (left), $\chi=[0.20-0.90]$ (middle) and $\chi=[0.20-0.24]$ (right) of a box simulation, compared with higher resolution and ${\cal C}=8$.}
	\label{fig:box_spectra_chis}
\end{figure*}

In Fig.~\ref{fig:box_spectra_chis} we compare the energy spectra obtained from simulations with two resolutions, $N=\{128,256\}$, for different metric backgrounds.  First, it is represented the spectra energy with $\chi=1$ (left), then a curved metric with a range of $\chi=[0.20-0.90]$ (middle), and finally an extreme case with $\chi=[0.20-0.24]$ (right panel). With $\chi=1$, the magnetic energy spectra for {\tt KH128C8} is above the case {\tt KH256C0}, as it was already shown previously. The same behavior, significantly moderated, is also observed when there is a large variation of the curved metric in the simulation domain (middle panel). In this case, the magnetic spectra for {\tt KH128C8} also rises with respect to {\tt KH128C0}, but it is only slightly above the {\tt KH256C0} case. Finally, when the metric curvature is really strong and homogeneous (right panel), turbulence is strongly suppressed, since the fluid can not move freely on such strong gravitational field. Therefore, all the curves in the magnetic energy spectra are comparable. The fact that the {\tt KH128C8} and {\tt KH256C0} cases also coincide here reinforces the idea that LES with the SGS gradient model is actually working as expected, not producing artificial enhancements of magnetic energy.

\begin{figure}

	\includegraphics[width=0.40\textwidth]{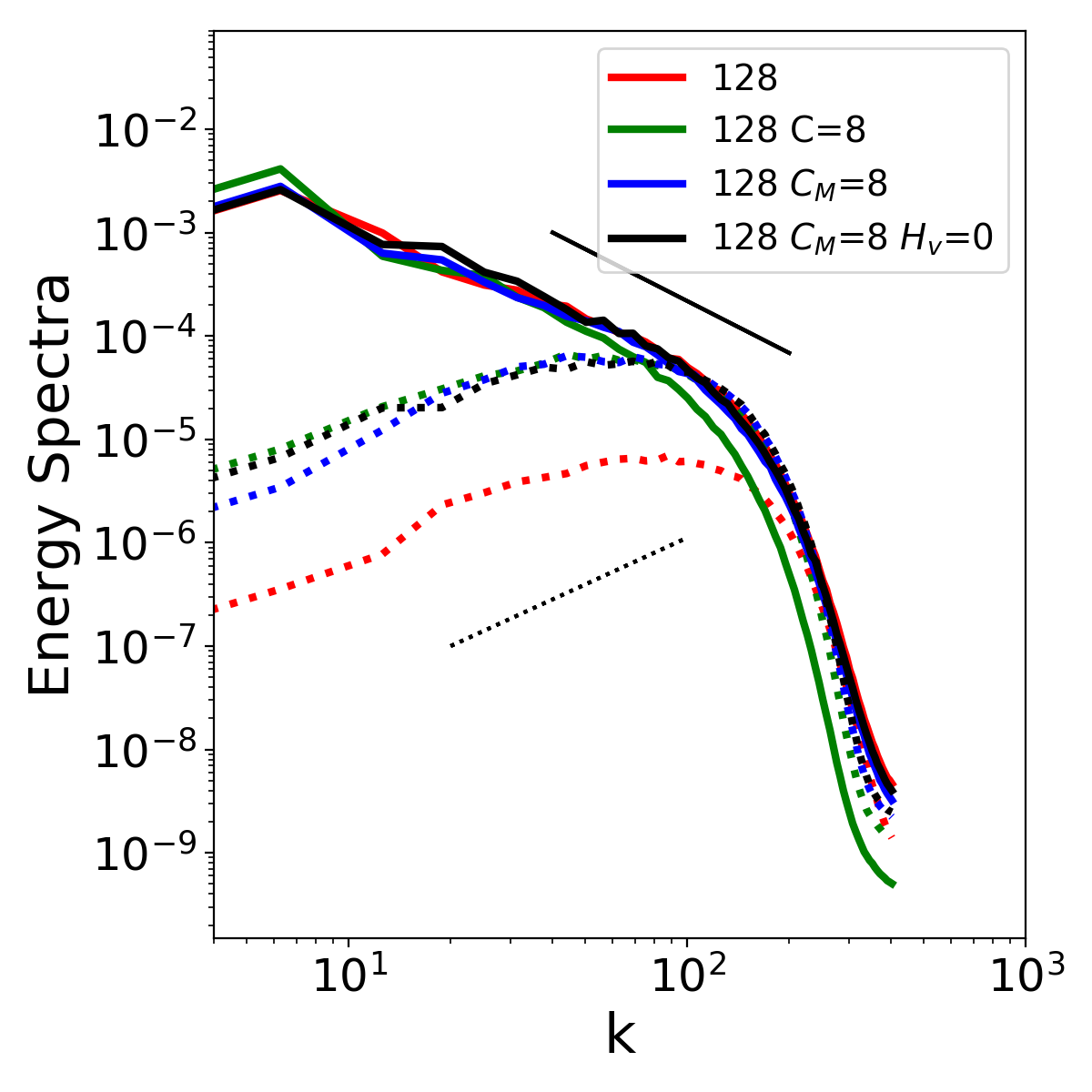}
	\caption{{\em Box simulations in 3D: relevant SGS terms for dynamo}. Kinetic (solid) and magnetic (dashed) energy spectra at $t=20$ comparing different cases: ${\cal C}=\{0,8\}$, all ${\cal C_X}=0$ except ${\cal C_M}=8$ and finally the latest one setting additionally $H_v=0$, such that the only SGS tensors activated are the ones appearing in Newtonian MHD. Clearly, these terms are dominating the small scale dynamics in this problem.  The black, thin solid line and dots represent the Kolmogorov $\propto k^{-5/3}$ and Kazantsev $\propto k^{3/2}$ slopes, respectively, as references.
	}
	\label{fig:box_spectra_vel_CM}
\end{figure}

Fig.~\ref{fig:box_spectra_vel_CM} displays the  energy spectra for the lowest resolution $N=128$ when only some of the SGS terms are activated. In particular, two new specific cases are explored: when all the coefficients ${\cal C}$ are set to zero except the one corresponding to ${\tau_M}$, and when in the previous case we additionally  set $H_v=0$. Clearly, almost all the rising of the magnetic energy spectra of {\tt KH128C8} is due to the Newtonian analogous of the magnetic SGS terms, achieved with $H_v=0$ and ${\cal C_M} \geq 1$. This means that this contribution, which corresponds to the nonrelativistic limit
\begin{equation}
H^{ki}_{M} \simeq 4 \, \nabla \overline{B}^{[i} \cdot \nabla \widetilde{v}^{k]} ~~,
\end{equation}
dominates over the other terms (at least for these physical conditions), indicating an easy way to model the feedback of SGS dynamo at small scales.

\section{Discussion}\label{sec:discussion}

In this paper, we have introduced the first GRMHD explicit LES with the physically-agnostic gradient SGS model. The aim is to better capturing small-scale effects of MHD turbulence at relevant astrophysical scenarios. Here we have focused on validating our numerical code, which combines our proposed SGS model for GRMHD with the use of high-order accurate numerical schemes, in box simulations developing the KHI with different parameters.

The extension from our previous studies of nonrelativistic \cite{vigano19b} and special relativistic \cite{carrasco19} MHD, to full GR, was carried under the following assumptions: (i) the SGS model is constructed starting from the filtering operation (associated to the numerical discretization) at the $3+1$ decomposition level; (ii) when filtering the MHD equations, the metric components are considered ``transparent'' to the gradient operators involved in the SGS tensors; and (iii) all SGS corrections arising from the Einstein equations are neglected. 
The assumption (i) means the construction of our model is not fully co-variant. Indeed, we argue that the need for a SGS model (to effectively capture part of the missing small scale dynamics) arises from the discretization itself, which is not an invariant operation. Thus, we find quite natural the SGS model adapts instead to the particular discretization at the $3+1$ level. It is worth to emphasize that the gradient model (by construction, as most SGS models) does not modify the continuous limit, nor the principal part of the evolution system. In this sense, the inclusion of the SGS terms is analogous to the numerical reconstruction methods, which also introduce a violation of the GR covariance.
The approximation (ii) has to do with the fact that the metric components can be considered smooth in comparison with the turbulent MHD fields, and thus, the contribution from their gradients in the SGS tensors should be sub-dominant. 
The last assumption (iii), linked to the previous one, states the dominant SGS corrections on the turbulent MHD variables are not expected to produce significant deviations on the spacetime geometry. We expect these assumptions to be fairy reasonable in the context of BNS mergers, where the large variations of the turbulent fields at small scales should contribute the dominant effects we aim to capture with the model.

We test our approach using 2D and 3D bounding box simulations of the relativistic KHI in a curved background, finding essentially the same results as in their nonrelativistic and special relativistic counterparts \cite{vigano19b, carrasco19} and supporting the present GR extension of our SGS model. We have considered a variety of different scenarios with different resolutions, and compared via \textit{a-priori} tests the SFS residuals within a certain scale range $[S_f \Delta,\Delta]$ with the proposed SGS model. We obtain best-fit values ${\cal C}\sim1-2$ (as expected) and high values ${\cal P}\gtrsim 0.8$ for $S_f=2$. This is consistent with our previous studies and indicate that the SGS is suitable to fit well the SFS residuals down to $S_f \lesssim 4$ at least (gradually degrading for smaller scales). The most important novelty of the \textit{a-priori} tests presented here are the inclusion in the problem of nontrivial metric components (i.e., curvature) and strong shocks in the fluid initial conditions.

Moreover, we have also performed LES with the proposed SGS models, thus allowing \textit{a-posteriori} tests, in which a high-resolution run is compared against low-resolution explicit LES with different values of  ${\cal C}\sim {\cal O}(1-10)$. The integrated magnetic energy in the KHI rises as one increases the resolution, while similar behaviour is shown to occur when increasing ${\cal C}$ at a lower resolution run including SGS terms. When spectra are looked, the $N=128$ run with ${\cal C}=8$ can achieve comparable magnetic energy spectrum with respect to high-resolution cases. Note that ${\cal C}$ significantly larger than 1 are needed. In order to explain this, note that in our previous work \cite{vigano19b}, we saw that including the next-leading order terms in the Taylor expansion of the SGS tensors give negligible differences in both magnetic energy and spectra, as compared with the first order approximation. We can conclude the required relatively high values of ${\cal C}\gtrsim 1$ can be associated to the intrinsic dissipation of the numerical finite-difference scheme employed (being likely smaller in case of using the intrinsically less dissipative spectral methods).

With the tests here presented, we have consistently extended to GRMHD the good performance of the gradient SGS model already found in nonrelativistic MHD \cite{muller02a,vigano19b} and special relativistic MHD \cite{carrasco19}. Overall, we have shown how the implementation of the gradient SGS terms in GRMHD LES can indeed reproduce part of the feedback given by unresolved dynamics over the large scales.

The main applicability relies on the fact that the mere inclusion of the gradient SGS terms allows the saving of at least one order of magnitude in computational time, providing similar numerical results.
This approach is then promising and supports our goal of applying the proposed gradient SGS model for GRMHD to realistic astrophysical scenarios where small scale dynamics can be crucial, like the supposedly spectacular growth of the magnetic field occurring during the merger of BNS, and the consequences it has for the production of observable jets.

\subsection*{Acknowledgments} 
We acknowledge support from the Spanish Ministry of Economy and Competitiveness grants FPA2013-41042-P and AYA2016-80289-P (AEI/FEDER, UE). CP also acknowledges support from the Spanish Ministry of Education and Science through a Ramon y Cajal grant. DV is supported by the ERC Consolidator Grant "MAGNESIA" (nr.~817661) and the Spanish grant SGR~2017-1383. 
We thankfully acknowledge the computer resources at MareNostrum and the technical support provided by Barcelona Supercomputing Center, with the time granted through the $17^{th}$ PRACE regular call (project Tier-0 GEEFBNSM, P.I. CP) and the RES calls AECT-2018-1-0005 (P.I. DV), AECT-2019-1-0007 (P.I. DV).

\bibliographystyle{utphys}
\bibliography{turbulence}
 
\end{document}